Research Article

Amanda K. Glazer* and Samuel D. Pimentel

# Robust inference for matching under rolling enrollment

**Abstract:** Matching in observational studies faces complications when units enroll in treatment on a rolling basis. While each treated unit has a specific time of entry into the study, control units each have many possible comparison, or "pseudo-treatment," times. Valid inference must account for correlations between repeated measures for a single unit, and researchers must decide how flexibly to match across time and units. We provide three important innovations. First, we introduce a new matched design, GroupMatch with instance replacement, allowing maximum flexibility in control selection. This new design searches over all possible comparison times for each treated-control pairing and is more amenable to analysis than past methods. Second, we propose a block bootstrap approach for inference in matched designs with rolling enrollment and demonstrate that it accounts properly for complex correlations across matched sets in our new design and several other contexts. Third, we develop a falsification test to detect violations of the timepoint agnosticism assumption, which is needed to permit flexible matching across time. We demonstrate the practical value of these tools via simulations and a case study of the impact of short-term injuries on batting performance in major league baseball.

**Keywords:** matching, block bootstrap, repeated measures, falsification test

**MSC 2020:** 62F40

## 1 Introduction

Matching methods attempt to estimate average causal effects by grouping each treated unit with one or more otherwise similar controls and using paired individuals to approximate the missing potential outcomes. Assuming that paired individuals are sufficiently similar on observed attributes and that no important unobserved attributes confound the comparison, the difference in outcomes approximates the impact of treatment for individuals in the pair [1]. Despite matching's transparency and intuitive appeal, it faces complications in datasets containing repeated measures for the same individuals over time. When only a single time of treatment is present, the primary challenge is deciding how to construct matching distances from pre-treatment repeated measures and assess outcomes using post-treatment repeated measures [2]. The situation is more complex under rolling enrollment, or staggered adoption, when individuals opt into treatment at different times [3]. Several authors [4–7] proceed by matching each treated unit to the version of the control unit present in the data at the time of treatment. For example, in Imai et al. [7]'s reanalysis of data from Acemoglu et al. [8] on the impact of democratization on economic growth, countries undergoing democratizing political reforms are matched to similar control countries not undergoing such reforms in the same year.

*** Corresponding author: Amanda K. Glazer,** Department of Statistics, University of California, Berkeley, 367 Evans Hall, Berkeley, CA 94720, USA, e-mail: amandaglazer@berkeley.edu
**Samuel D. Pimentel:** Department of Statistics, University of California, Berkeley, 367 Evans Hall, Berkeley, CA 94720, USA

Although this method is logical whenever strong time trends are present, in other cases, it may overemphasize similarity on time at the expense of other variables. Bohl et al. [9] study the impact of serious falls on subsequent healthcare expenditures for elderly adults using patient data from a large healthcare system. While patients who fall could be matched to patients who appear similar based on recent health history on the calendar date of the fall, the degree of similarity in health histories is likely much more important than the similarity of the exact date at which each patient is measured. Following this idea, the GroupMatch algorithm [10] constructs matches optimally across time, prioritizing matching on important covariates over ensuring that units are compared at the same point in time.

Another example where rolling enrollment arises is in major league baseball (MLB). Quantifying the impact of injury on player performance in professional sports is important for both managers and players themselves. Increasingly, players are valued and compensated in a manner driven by quantitative metrics of past performance, but injuries have potential to disrupt the continuity between past and future performance [11–14]. One way to quantify impact in this setting is by finding the difference between the value of a performance metric the player would have achieved in the absence of injury and the value of the same metric achieved after a given injury. This quantity can be estimated using matching. However, players do not all get injured at the same time, so GroupMatch is a natural fit here. It allows us to match injured and noninjured players flexibly across time, because we likely do not care whether a player injured on June 1 is matched to a noninjured player on exactly the same day or a couple weeks earlier, e.g., May 15, so long as those players are sufficiently similar on other covariates such as recent performance.

Several challenges remain outstanding for matching methods under rolling enrollment. GroupMatch's flexible approach relies heavily on a strong assumption that time itself is not a confounder, and discussion of checking this assumption has been minimal so far. Even when flexible matching is warranted, the presence of multiple copies of the same control individual necessitates a constraint to ensure that a treated unit is not simply paired to multiple slightly different copies of the same control; several choices of this constraint exist permitting varying degrees of flexibility, and users must choose among them. Most importantly, for both GroupMatch and methods that match exactly in time, there is substantial ambiguity about how to conduct valid inference. When multiple copies of a control individual are forbidden from appearing in the matched design, randomization inference may be used [5,10], but no strong guarantees exist outside this special case.

In what follows, we present several innovations that greatly enhance the toolkit for matching and treatment effect evaluation under rolling enrollment. First, we introduce a new matched design called GroupMatch with instance replacement, which has computational, analytical, and statistical advantages over existing designs in many common settings. Second, we give a comprehensive characterization of a new block-bootstrap-based method for inference that applies broadly across existing methods for matching under rolling enrollment, including our new design. The block-bootstrap approach was originally suggested by Imai et al. [7] and is based on related work in the cross-sectional case by Otsu and Rai [15], but until now has not carried any formal guarantee. Finally, we introduce a falsification test to provide a partial check of the assumption of timepoint agnosticism underpinning GroupMatch's validity, empowering investigators to extract evidence from the data about this key assumption before matching. We prove the validity of our bootstrap method under the most relevant set of constraints on reuse of controls, and we also demonstrate the effectiveness of both the falsification test and the bootstrap inference approach through simulations and an analysis of injury data in MLB. In particular, the bootstrap method shows improved performance over linear regression-based approaches to inference often applied in similar settings.

This article is organized as follows. Section 2 presents the basic statistical framework and reviews the GroupMatch framework, inference approaches for matching designs, and other related literature. In Section 3, we introduce a new constraint for use of controls in GroupMatch designs, leading to a new design called GroupMatch with instance replacement. Section 4 presents a block bootstrap inference approach for matching under rolling enrollment, and Section 5 evaluates it via simulation. In Section 6, we present a falsification test for the assumption that time is not a confounder. In Section 7, we apply our methods to evaluate whether minor injuries impact short-term MLB performance. Section 8 concludes this article.

# 2 Statistical framework

## 2.1 Setting and notation

We observe $n$ subjects. For each subject $i$ in the study, we observe repeated measures $(Y_{i,t}, \mathbf{X}_{i,t})$ for time-points $t = 1, \dots, T$, where $Y_{i,t}$ is an outcome of interest and $\mathbf{X}_{i,t}$ is a vector of covariates. We also observe a time of treatment initiation $T_i$ for each subject, with $T_i \in \{1, \dots, T\}$ for subjects who receive treatment at some point and $T_i = \infty$ for those who remain controls at all observed timepoints. We specify a burn-in period of length $L - 1$ during which no individuals are treated, i.e., $T_i \geq L$ for all $i$ (or allow treatment at $t = 1$ by setting $L = 1$). We denote the collection of repeated measures for each subject $i$, along with $T_i$, as the trajectory $O_i$.

For clarity, we focus on "instantaneous" effects of treatment, with outcomes measured immediately following treatment. Let $Y_{i,t}(0)$ be the potential outcome for unit $i$ that would have been observed at time $t$ if $T_i > t$, and let $Y_{i,t}(1)$ be the potential outcome that would have been observed if $T_i = t$. The finite sample average effect of treatment on the treated (ATT) is denoted by $\Delta$:

$$\Delta = \frac{1}{N_1}\sum_{i=1}^{N}\sum_{t=1}^{T}1\{t = T_i\}[Y_{i,t}(1) - Y_{i,t}(0)] = \frac{1}{N_1}\sum_{i=1}^{N}D_i[Y_{i,t=T_i}(1) - Y_{i,t=T_i}(0)].$$

Here, the $D_i$ variable is introduced as a convenient shorthand to indicate whether $T_i < \infty$. We assume that trajectories $O_i$ are sampled independently from some infinite population, although we do not assume independence of observations within the same trajectory. By defining expectation $E(\cdot)$ with respect to sampling from this population, we define the population ATT as $\Delta_{\text{pop}} = E(\Delta)$. For future convenience, we also introduce a concise notation for conditional expectation (again, over the sampling distribution) of potential outcomes given no treatment through time $t$ and the covariates observed in the previous $L$ timepoints:

$$\mu_0^t(\mathbf{X}) = E[Y_{i,t}(0) | \{X_{i,t'}\}_{t'=t-L+1}^{t'=t} = \mathbf{X}, T_i > t].$$

Throughout, we relax notation slightly by writing $\mu_0(\mathbf{X}_{i,t})$ to indicate conditional expectation given the $L$ lagged values of $\mathbf{X}_i$ directly preceding time $t$, inclusive.

The potential outcomes framework adopted here represents one of many possible framings for studies with rolling enrollment. Pimentel et al. [10] define potential outcomes as functions of the length of time since treatment initiation, while both Ben-Michael et al. [3] and Athey and Imbens [16] define them as functions of the time of treatment initiation for the subject in question. In principle, these alternate constructions are much richer than ours, allowing for much more general and complicated patterns of effects, but in practice, all these authors use simplifying assumptions or focus on estimands that reduce attention to at most two potential outcomes of interest for each individual at each timepoint. For example, both Pimentel et al. [10] and Ben-Michael et al. [3] allow for treatment effects to be measured at some follow-up time postdating the time of treatment rather than focusing on instantaneous effects, but since the length of follow-up is fixed in advance, only two potential outcomes ever need to be considered for each unit. Similarly, the "no anticipation" assumption of both Ben-Michael et al. [3] and Athey and Imbens [16] ensures that there is exactly one potential outcome of interest associated with the control condition for each individual, and the "invariance to history" assumption of Athey and Imbens [16] collapses distinctions among potential outcomes under treatment. As such, the results we present in the following sections extend easily to all the potential outcomes frameworks just described, making the appropriate substitutions for our $Y_{i,t}(1)$ and $Y_{i,t}(0)$. While the potential outcomes framework of Imai et al. [7] is much more general than all the previously mentioned works in allowing subjects to revert from treatment back to control, our framing also extends easily to it in the special case when no exit from treatment is allowed.

## 2.2 Identification assumptions

Pimentel et al. [10] studied the following difference-in-means estimator in designs where each treated unit is matched to $C$ control observations. $M_{it,jt'}$ is an indicator for whether subject $i$ at time $t$ has been matched to subject $j$ at time $t'$:

$$\hat{\Delta} = \frac{1}{N_1}\sum_{i=1}^{n} D_i \left[ Y_{i,t=T_i} - \frac{1}{C}\sum_{j=1}^{N}\sum_{t'=1}^{T} M_{iT_i,jt'} Y_{j,t'} \right].$$

Pimentel et al. [10] show that this estimator is unbiased for the population ATT under the following conditions:

(1) Exact matching: matched units share identical values for covariates in the $L$ timepoints preceding treatment.

(2) $L$-ignorability: conditional on the covariate history over the previous $L$ timepoints and the absence of treatment before baseline, an individual's potential outcome at a given time is independent of the individual's overall treatment status. Formally,

$$\{T_i < \infty\} \perp\!\!\!\perp Y_{i,t}(0) | T_i > t-1, \{X_{i,s}\}_{s=t-L+1}^{t}, \quad \forall i.$$

Intuitively, this assumption prevents unobserved confounding that makes potential outcomes for treated subjects systematically different from those that remain controls even after accounting for information from a baseline period.

(3) Timepoint agnosticism: mean potential outcomes under control do not differ for any instances with identical covariate histories at different timepoints. Formally, for any set of $L$ covariate values $\mathbf{X}$,

$$\mu_0^t(\mathbf{X}) = \mu_0^{t'}(\mathbf{X}) = \mu_0(\mathbf{X}) \quad \text{for any } 1 \le t, t' \le T.$$

This assumption ensures that matching across time is reasonable by ruling out time trends other than those captured by time-varying covariates. For clarity, we drop the $t$ superscript when discussing the conditional expectation $\mu_0(\mathbf{X})$ in what follows, with the exception of Section 6 where we temporarily consider failures of this assumption.

(4) Covariate $L$-exogeneity: future covariates do not encode information about the potential outcome at time $t$ given covariates and absence of treatment over the previous $L$ time points. Formally,

$$(X_{i,1}, \ldots, X_{i,T}) \perp\!\!\!\perp Y_{i,t}(0) | T_i > t-1, \{X_{i,s}\}_{s=t-L+1}^{t}, \quad \forall i.$$

Like time agnosticism, covariate $L$-exogeneity is important to justify considering past and future instances from a control trajectory as part of the matching procedure. If future instances' covariates include or are correlated with past instances' outcomes, then we may indirectly match on study outcomes introducing bias into our estimation step [17,18]. Covariate $L$-exogeneity ensures that future covariates are safe to consider during the design stage.

(5) Overlap: given that a unit is not yet treated at time $t-1 \ge L$, the probability of entering treatment at the next time point is neither 0 nor 1 for any choice of covariates over the $L$ timepoints at and preceding $t$.

$$0 < P(T_i = t | T_i > t-1, X_i^t, \ldots, X_i^{t-L+1}) < 1 \quad \forall t > L.$$

While this assumption is not stated explicitly in the study by Pimentel et al. [10], we note that the authors rely on an overlap assumption of this type in the proof of their main result.

The exact matching assumption is no longer needed for asymptotic identification of the population ATT if we modify the estimator by adding in a bias correction term. As shown in the studies by Otsu and Rai [15] and Abadie and Imbens [19], we first estimate the conditional mean function $\mu_0(\mathbf{X})$ of the potential outcomes and use this outcome regression to adjust each matched pair for residual differences in covariates not addressed by matching. As outlined in the study by Abadie and Imbens [19], bias correction leads to asymptotic consistency under regularity conditions on the potential outcome mean estimator $\hat{\mu}(\cdot)$ (for further discussion of regularity assumptions on $\widehat{\mu}_0(\cdot)$ see the proof of Theorem 1 in Section A of the supplemental appendix). Many authors have also documented benefits from adjusting matched designs using outcome models [20,21]. The specific form of our bias-corrected (i.e., model-adjusted) estimator is as follows:

$$\hat{\Delta}_{\text{adj}} = \frac{1}{N_1}\sum_{i=1}^{n} D_i\left[\left(Y_{i,t=T_i} - \hat{\mu}_0\left(\mathbf{X}_{i,T_i}\right)\right) - \frac{1}{C}\sum_{j=1}^{N}\sum_{t'=1}^{T} M_{iT_i,jt'}(Y_{j,t'} - \hat{\mu}_0(\mathbf{X}_{i,t'}))\right].$$

Datasets with many variables, especially continuous variables or variables with many categories, ensure that exact matching is rarely possible in practice, and in light of this, we focus primarily on estimator $\widehat{\Delta}_{\text{adj}}$ in what follows.

As discussed by Imai et al. [7] for settings without rolling enrollment, identification is possible under weaker assumptions if a difference-in-differences estimator is used instead of the difference-in-means. While we focus primarily on the simpler bias-corrected difference-in-means estimator for clarity of exposition, the difference-in-differences approach also offers advantages for our setting, and the new matching and inference strategies we propose extend naturally to such estimators. We provide further discussion in Section 4.3.

# 3 GroupMatch with instance replacement

Before discussing our method for inference in general matched designs under rolling enrollment, we introduce a new type of GroupMatch design. Pimentel et al. [10] described two different designs produced by GroupMatch and denoted them Problems A and B. For increased clarity, we refer to Problem A as GroupMatch without replacement, and Problem B as GroupMatch with trajectory replacement.

(1) **GroupMatch without replacement**: Each control unit can be matched to at most one treated unit. If a treated unit is matched to an instance of a control unit, no other treated unit can match to (any instance of) that control unit.
(2) **GroupMatch with trajectory replacement**: Each control *instance* can be matched to at most one treated unit. Each treated unit can match to no more than one instance from the same control trajectory. However, different treated units can match to different instances of the same control trajectory, so a single control trajectory can contribute multiple distinct instances to the design.

As our chosen names for these designs suggest, their relative costs and benefits reflect the choice between matching without and with replacement in cross-sectional settings. As discussed by Hansen [22], matching without replacement (in which each control may appear in at most one matched set), leads to less similar matches compared to matching with replacement (in which controls can reappear in many matched sets) since in cases where two treated units share the same nearest control, only one can use it. On the other hand, matching without replacement frequently leads to estimators with lower variance than those from matching with replacement, where an individual control unit may appear in many matched sets, making the estimator more sensitive to random fluctuations in its response. Thus, one aspect of choosing between these designs is a choice about how to strike a bias-variance tradeoff. The other important aspect distinguishing these designs is that randomization inference, which is based on permuting treatment assignments in each matched set independently of others, generally requires matching without replacement. Specifically, when multiple controls may be matched to each treated unit and replacement is allowed, the resulting configuration of treated and control units no longer resembles the design of a blocked or matched experiment.

These same dynamics play out in comparing GroupMatch without replacement and GroupMatch with trajectory replacement. GroupMatch without replacement ensures that responses in distinct matched sets are statistically independent (under a model in which trajectories are sampled independently), allowing for randomization inference, and ensures that the total weight on observations from any one control trajectory can sum only to $1/C$, ensuring that the estimator's variance cannot be too highly inflated by a single trajectory a large weight. The resulting data configuration also resembles what might be obtained in a sequential experimental design employing matching-on-the-fly as discussed by Kapelner and Krieger [23] and Pimentel et al. [10], and inference may be conducted using the associated randomization distribution. On the other hand, GroupMatch with trajectory replacement leads to higher-quality matches and reduced bias in matched pairs, although overlap among the matched sets formed makes it difficult to envision a corresponding "target trial."

We suggest a third GroupMatch design that leans even further towards expanding the potential control pool and reducing bias.

(3) **GroupMatch with instance replacement**: Each treated unit can match to no more than one instance from the same control unit, but control instances can be matched to more than one treated unit.

GroupMatch with instance replacement is identical to GroupMatch without trajectory replacement except that it allows repetition of individual instances within the matched design as well as nonidentical instances from the same trajectory. As such, it is guaranteed to produce higher-quality matches than GroupMatch without trajectory replacement, but may lead to higher-variance estimators since individual instances may receive weights larger than $1/C$. Figure 1 illustrates these three GroupMatch methods with a toy example that matches injured baseball players to noninjured players based on on-base percentage (OBP).

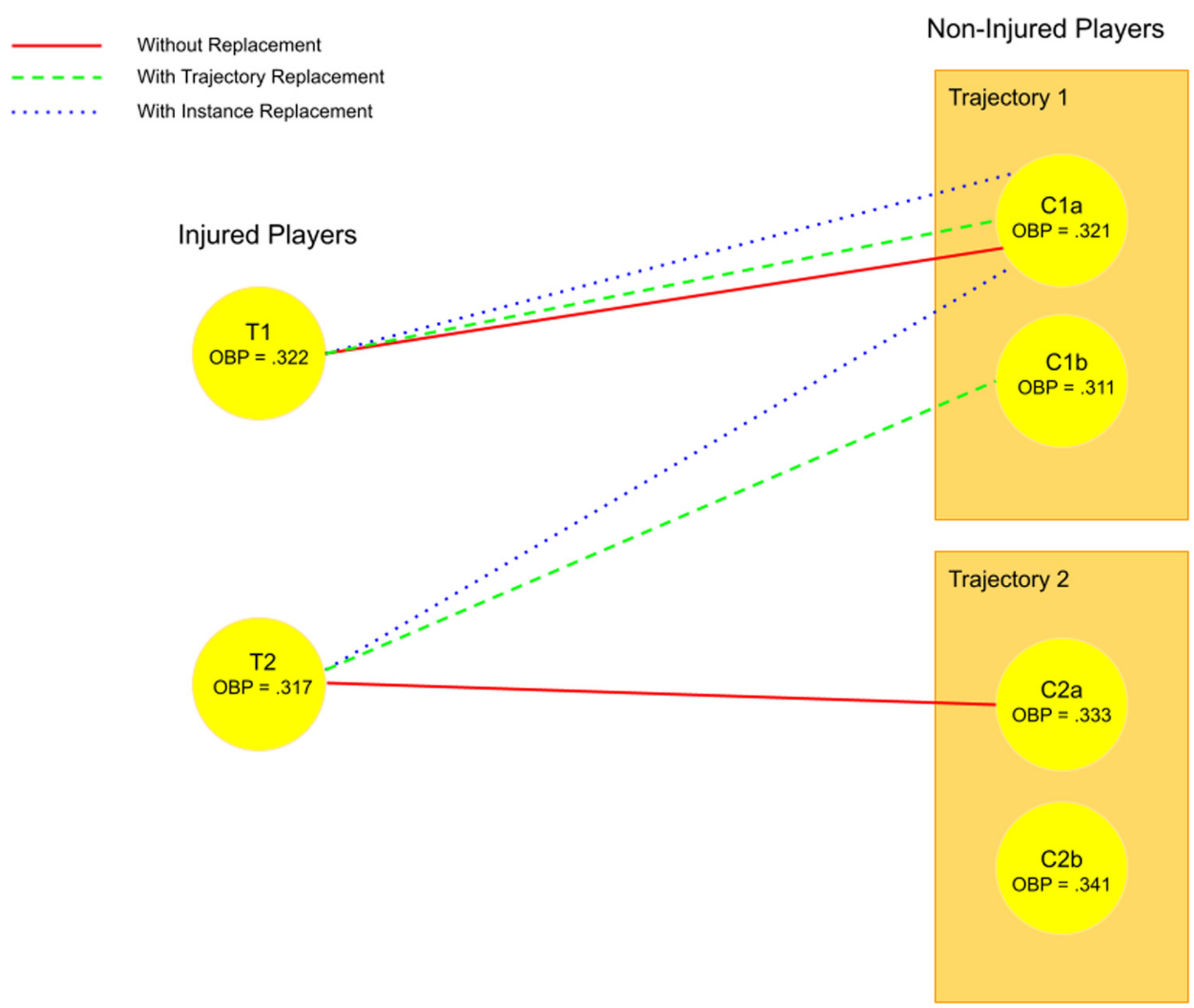

**Figure 1:** Toy example illustrating the three GroupMatch matching methods. Two injured baseball players (T1 and T2) are matched 1-1 to noninjured baseball players (C1a/b and C2a/b) based on player OBP. Each noninjured player has two pseudo-injury times or instances. Under GroupMatch without replacement, T2 must match to an instance in Trajectory 2 because at most one instance from Trajectory 1 can participate in the match. Under GroupMatch with trajectory replacement, T2 can match to C1b but not to C1a, since multiple control instances can be chosen from the same trajectory as long as they are distinct. Under GroupMatch with instance replacement, both T1 and T2 are able to match to C1a. However, if each treated instance were matched to two control instances instead of one, GroupMatch with instance replacement would still forbid either T1 or T2 to match to a second instance in Trajectory 1.

In practice, we view GroupMatch with instance replacement as a more attractive approach than GroupMatch with trajectory replacement almost without exception. One reason is that while the true variance of estimators from GroupMatch with instance replacement may often exceed that of estimators from GroupMatch with trajectory replacement by a small amount, our recommended approach for *estimating* the variance and conducting inference are not able to capture this difference. As we describe in Section 4, in the absence of a specific parametric model for correlations within a trajectory, inference proceeds in a conservative manner by assuming arbitrarily high correlations within a trajectory (much like the clustered standard error adjustment in linear regression). Since the variance advantage for GroupMatch with trajectory replacement arises only when correlations between instances within a trajectory are lower than one, the estimation strategy is not able to take advantage of them. This disconnect means that GroupMatch with trajectory replacement will not generally lead to narrower empirical confidence intervals, much as variance gains associated with paired randomized trials relative to less-finely stratified randomized trials may not translate into reduced variance estimates [24].

A second important advantage of GroupMatch with instance replacement is its computational and analytical tractability relative to the other GroupMatch designs. One way to implement GroupMatch with instance replacement as a network flow optimization problem is to remove a set of constraints in Pimentel et al. [10]'s Network B (specifically the upper capacity on the directed edges connected to the sink node), and in Sections 5 and 7, we use this implementation for its convenient leveraging of the existing `groupmatch` package in R. However, much more computationally efficient algorithms are also possible. Crucially, the removal of the constraint forbidding instance replacement means that matches can be calculated for each treated instance without reference to the choices made for other treated units; the $C$ best matches for a given treated unit are simply the $C$ nearest neighbor instances such that no two such control instances within the matched set come from the same trajectory. In principle, this allows for complete parallelization of the matching routine. On the analytical side, this aspect of the design makes it possible to characterize the matching algorithm as a generalized form of nearest neighbor matching, a strategy we adopt in the proof of Theorem 1 to leverage proof techniques used by Abadie and Imbens [25] for cross-sectional nearest neighbor matching. In light of these considerations, we focus primarily on GroupMatch with instance replacement in what follows, although the methods derived appear to perform well empirically for other GroupMatch designs too.

# 4 Block bootstrap inference

## 4.1 Inference methods for matched designs

Broadly speaking, there are two schools of thought in conducting inference for matched designs. One approach, spearheaded by Abadie and Imbens [19,25–27], views the raw data as samples from an infinite population and demonstrates that estimators based on matched designs (which in this framework are considered to be random variables, as functions of random data) are asymptotically normal. Inferences are based on the asymptotic distributions of matched estimators. A second approach, described in detail in Rosenbaum [28,29] and Fogarty [30], adopts the perspective of randomization inference in controlled experiments. Conditional on the structure of the match and the potential outcomes, the null distribution of a test statistic over all possible values of the treatment vector is obtained by permuting values of treatment within matched sets. When matches are exact and unobserved confounding is absent, strong finite sample guarantees hold for testing sharp null hypotheses without further assumptions on outcome variables. Asymptotic guarantees for weak null hypotheses may be obtained too, assuming a sequence of successively larger finite populations [31]. Well-developed methods of sensitivity analysis are also available.

As described in the study by Pimentel et al. [10], while standard methods of inference may be applied to GroupMatch without replacement, in which control individuals contribute at most one unit to any part of the match, none have been adequately developed for GroupMatch with trajectory replacement, in which distinct matched sets may contain different versions of the same control individual. For randomization inference, the barrier appears to be quite fundamental, because permutations of treatment within one matched set can no longer be considered independently for different matched sets. In GroupMatch with trajectory replacement, a treated unit receives treatment at one time and appears in a match only once; if treatment is permuted among members of a matched set so that a former control now attains treatment status, what is to be done about other versions of this control unit that are present in distinct matched sets? We note that similar issues arise when contemplating randomization inference for general cross-sectional matching designs with replacement, and we are aware of no solutions for randomization inference even in this simpler case.

In contrast, the primary issue in applying sampling-based inference to GroupMatch designs with trajectory replacement is the unknown correlation structure for repeated measures from a single control individual. The literature on matching with replacement provides estimators for pairs that are fully independent [27] and for cases in which a single observation appears identically in multiple pairs [25], but not for the intermediate case of GroupMatch with trajectory replacement where distinct but correlated observations appear in distinct matched sets. These issues extend beyond the GroupMatch family to any matched design under rolling enrollment in which control trajectories contribute to multiple matched sets, including those of Witman et al. [6] and Imai et al. [7].

In what follows, we give formal guarantees for a sampling-based inference method appropriate for general matching designs under rolling enrollment suggested by Imai et al. [7], which generalizes a recent proposal of Otsu and Rai [15] for valid sampling-based inference of cross-sectional matched studies using the bootstrap. Although the bootstrap often works well for matched designs without replacement [32], naive applications of the bootstrap in matched designs with replacement have been shown to produce incorrect inferences as a consequence of the failure of certain regularity conditions [26]. Intuitively, if matching is performed after bootstrapping the original data, multiple copies of a treated unit will necessarily all match to the same control unit, creating a clumping effect not present in the original data. However, Otsu and Rai [15] arrived at an asymptotically valid bootstrap inference method for matching by bootstrapping weighted and bias-corrected functions of the original observations *after* matching rather than repeatedly matching from scratch in new bootstrap samples. We show that a similar bootstrap approach, applied to entire trajectories of repeated measures in a form of the block bootstrap, provides valid inference for matched designs under rolling enrollment. Note that in our formal results, we focus on GroupMatch with instance replacement as the most difficult case, since the designs of Witman et al. [6] and Imai et al. [7] may be understood as restricted special cases in which matching on time is exact.

## 4.2 Block bootstrap

To conduct inference under GroupMatch with trajectory or instance replacement, we propose a weighted block bootstrap approach. We rearrange the GroupMatch ATT estimator from Section 2 as follows, letting $K_M(i,t)$ be the number of times the instance at trajectory $i$ and time $t$ is used as a match.

$$\begin{aligned}\hat{\Delta}_{\text{adj}} &= \frac{1}{N_1}\sum_{i=1}^{N} D_i\left[\left(Y_{i,T_i} - \hat{\mu}_0\left(\mathbf{X}_{i,T_i}\right)\right) - \frac{1}{C}\sum_{j=1}^{N}\sum_{t'=1}^{T} M_{iT_i,jt'}(Y_{j,t'} - \hat{\mu}_0(\mathbf{X}_{i,t'}))\right]\\ &= \frac{1}{N_1}\sum_{i=1}^{N}\left[D_i\left(Y_{i,T_i} - \hat{\mu}_0\left(\mathbf{X}_{i,T_i}\right)\right) - (1 - D_i)\sum_{t=1}^{T}\frac{K_M(i,t)}{C}(Y_{i,t} - \hat{\mu}_0(\mathbf{X}_{i,t}))\right] = \frac{1}{N_1}\sum_{i=1}^{N}\hat{\Delta}_i.\end{aligned}$$

Because different instances of the same control unit are correlated, we resample the *trajectory*-level quantities $\widehat{\Delta}_i$ rather than the instance-level quantities. Since the $\widehat{\Delta}_i$ are functions of the $K_M(i, t)$ weights in the original match, we do not repeat the matching process within bootstrap samples. In particular, we proceed as follows:

(1) Fit an outcome regression $\widehat{\mu}_0(\cdot)$ based on covariates in the previous $L$ timepoints using only control trajectories.
(2) Match treated instances to control instances using GroupMatch with instance replacement. Calculate matching weights $K_M(i, t)$ equal to the number of times the instance at time $t$ in trajectory $i$ appears in the matched design.
(3) Calculate the model-adjusted ATT estimator $\widehat{\Delta}_{\text{adj}}$.
(4) Repeat $B$ times:
   (a) Randomly sample $N$ elements $\widehat{\Delta}^*$ with replacement from $\{\widehat{\Delta}_1, \ldots, \widehat{\Delta}_N\}$.
   (b) Calculate the bootstrap bias-corrected ATT estimator $\widehat{\Delta}^*_{\text{adj}}$ for this sample of trajectories as follows:
$$\widehat{\Delta}^*_{\text{adj}} = \frac{1}{N_1}\sum_{i=1}^{N}\widehat{\Delta}^*_i.$$
(5) Construct a $(1-\alpha)$ confidence interval based on the $\alpha/2$ and $1-\alpha/2$ percentile of the $\widehat{\Delta}^*_{\text{adj}}$-values calculated from the bootstrap samples.

This method is essentially a block bootstrap, very similar to the method proposed in Imai et al. [7]. Note that while the recipe given earlier uses the nonparametric bootstrap, it may easily be generalized to other approaches such as the wild bootstrap and the Bayesian bootstrap. In particular, consider rewriting $\widehat{\Delta}^*_{\text{adj}}$ in terms of a new set of random variables $W_1^*, \ldots, W_N^*$ that we denote the bootstrap weights:
$$\widehat{\Delta}^*_{\text{adj}} = \frac{1}{N_1}\sum_{i=1}^{N}\widehat{\Delta}^*_i = \sum_{i=1}^{N}\frac{W_i^*}{\sqrt{N_1}}\widehat{\Delta}_i. \tag{1}$$
To recover the nonparametric bootstrap, the bootstrap weights $W_i^*$ are chosen as $Q_i/\sqrt{N_1}$, where $Q_i$ is the number of times subject $i$ is selected when sampling with replacement; if the $W_i^*$ are chosen instead by sampling from a scaled Dirichlet or scaled two-point distribution, we obtain the Bayesian bootstrap and the wild bootstrap, respectively (see Otsu and Rai [15] for specifics). To adapt the step-by-step algorithm for these approaches, we draw $W_i^*$s rather than $\widehat{\Delta}^*$s in step 4(a) and use (1) to calculate $\widehat{\Delta}^*_{\text{adj}}$ in step 4b.

Our main result given here shows the asymptotic validity of this approach. Several assumptions, in addition to Assumptions 2–5 in Section 2.2, are needed to prove this result. We summarize these assumptions verbally here, deferring formal mathematical statements to Section A.1 of the supplemental appendix. First, we require the covariates $X_i$ to be continuous with compact and convex support and a density both bounded and bounded away from zero. Second, we require that the conditional mean functions are smooth in $\mathbf{X}$, with bounded fourth moments. In addition, we require that conditional variances of the treated potential outcomes and conditional variances of nontrivial linear combinations of control potential outcomes from the same trajectory are smooth and bounded away from zero. We also require that conditional fourth moments of potential outcomes under treatment and linear combinations of potential outcomes under control are uniformly bounded in the support of the covariates. We also make additional assumptions related to the conditional outcome mean estimator $\widehat{\mu}_0(\cdot)$, specifically that the $kL$th derivative of the true conditional mean functions $\mu_1^t(\cdot)$ and $\mu_0(\cdot)$ exist and have finite suprema, and that the $\widehat{\mu}(\cdot)$ converges to $\mu_0(\cdot)$ at a sufficiently fast rate. Finally, we require mild regularity conditions on the bootstrap weights $W_i^*$, easily satisfied by construction in the bootstrap approaches we have mentioned. To state the theorem, we also define
$$\sqrt{N_1}\,U^* = \frac{1}{\sqrt{N_1}}\sum_{i=1}^{N}(\widehat{\Delta}^*_i - D_i\widehat{\Delta}_{\text{adj}}) = \sum_{i=1}^{N}W_i^*(\widehat{\Delta}_i - D_i\widehat{\Delta}_{\text{adj}}).$$

**Theorem 1.** *Under assumptions M, W, and R presented in Section A.1 of the supplemental appendix,*

$$\sup_r |\Pr\{\sqrt{N_1}U^* \le r|(\mathbf{Y},\mathbf{D},\mathbf{X})\} - \Pr\{\sqrt{N_1}(\hat{\Delta}_{\text{adj}} - \Delta) \le r\}| \xrightarrow{p} 0$$

*as* $N \to \infty$ *with fixed control:treated ratio C.*

Our regularity assumptions on the data-generating process and the regression estimator $\hat{\mu}_0(\cdot)$ are modeled closely on those of Abadie and Imbens [25] and later Otsu and Rai [15], and our proof technique is very similar to arguments in Otsu and Rai [15]. Briefly, $U^*$ is decomposed into three terms that correspond to deviations of the potential outcome variables around their conditional means, approximation errors for $\hat{\mu}_0(\mathbf{X})$ terms as estimates of $\mu_0(\mathbf{X})$ terms, and deviations of conditional average treatment effects $\mu_1^t(\mathbf{X}) - \mu_0(\mathbf{X})$ around the population ATT $\Delta$. Regularity conditions on the data-generating process ensure that the conditional average treatment effects converge quickly to the population ATT. Regularity assumptions on the regression estimator, combined with bounds on the largest nearest-neighbor discrepancies in $\mathbf{X}$ vectors due originally to Abadie and Imbens [25] and adapted to the GroupMatch with instance replacement design, show that the deviation between $\hat{\mu}_0(\cdot)$ and $\mu_0(\cdot)$ disappears at a fast rate. Finally, a central limit theorem applies to the deviations of the potential outcomes. For details, see Section A of the supplemental appendix.

## 4.3 Difference-in-differences estimator

While we have focused so far on the difference-in-means estimator, Imai et al. [7] recommend a difference-in-differences estimator for matched designs with rolling enrollment in the context of designs that match exactly on time. This estimator can be used under rolling enrollment as well, taking the following form under bias correction:

$$\begin{aligned}\hat{\Delta}_{DiD} &= \frac{1}{N_1}\sum_{i=1}^{N} D_i\Bigg[\Big(\big(Y_{i,T_i} - \hat{\mu}_0(\mathbf{X}_{i,T_i})\big) - \big(Y_{i,T_i-1} - \hat{\mu}_0(\mathbf{X}_{i,T_i-1})\big)\Big) - \frac{1}{C}\sum_{j=1}^{N}\sum_{t'=1}^{T} M_{iT_i,jt'}((Y_{j,t'} - \hat{\mu}_0(\mathbf{X}_{i,t'})) \\ &\qquad - (Y_{j,t'-1} - \hat{\mu}_0(\mathbf{X}_{i,t'-1})))\Bigg] \\ &= \frac{1}{N_1}\sum_{i=1}^{N} D_i\Bigg[\Big(\big(Y_{i,t=T_i} - \hat{\mu}_0(\mathbf{X}_{i,T_i})\big) - \big(Y_{i,t=T_i-1} - \hat{\mu}_0(\mathbf{X}_{i,T_i-1})\big)\Big) - (1 - D_i)\sum_{t=1}^{T}\frac{K_M(i,t)}{C}((Y_{i,t} - \hat{\mu}_0(\mathbf{X}_{i,t})) \\ &\qquad - (Y_{i,t-1} - \hat{\mu}_0(\mathbf{X}_{i,t-1})))\Bigg] = \frac{1}{N_1}\sum_{i=1}^{N}\hat{\Delta}_i^{DiD}.\end{aligned}$$

This estimator requires $L$ lags to be measured at time $T_i - 1$, so a burn-in period of length $L$ rather than $L - 1$ is needed.

A key advantage of this bias-corrected differences-in-differences estimator is that it relies on different identification assumptions than the bias-corrected difference-in-means: essentially, any assumption previously made on the potential outcome $Y_{i,t}(0)$ must now hold instead only for the post–pre potential outcome difference $Y_{i,t}(0) - Y_{i,t-1}(0)$. The resulting assumptions tend to be substantively weaker. In particular, Imai et al. [7] highlight how the $L$-ignorability assumption can be replaced by a parallel trends assumption that requires only that post–pre differences in potential outcomes be conditionally independent of treatment, allowing for different unobserved outcome intercepts for different individuals. The time-agnosticism assumption also becomes weaker when formulated for outcome differences, allowing for a constant linear trend in potential outcome means rather than requiring them to be invariant to time conditional on covariates.

We can easily adapt the results of the previous section to show that the block bootstrap gives valid inference for the difference-in-difference estimator when these identification assumptions hold. The inference procedure simply requires bootstrapping the $\hat{\Delta}_i^{DiD}$ terms in place of the $\hat{\Delta}_i$s defined earlier. While the regularity conditions given for Theorem 1 suffice for the new estimator, the proof (as presented in Section A of the supplemental appendix) requires mild modification to work for this difference-in-differences estimator. In particular, the variance formulas include additional covariance terms. For more details, see Section A.3 of the supplemental appendix.

# 5 Simulations

We now explore the performance of weighted block bootstrap inference via simulation. In particular, we investigate coverage and length of confidence intervals compared to those obtained by conducting parametric inference for weighted least squares (WLS) estimators with and without cluster-robust error adjustment for controls from the same trajectory.

## 5.1 Data generation

We generate eight covariates, four of them uniform across time for each individual $i$, (i.e., they take on the same value at every timepoint):

$$\begin{pmatrix} X_{3,i} \\ X_{4,i} \end{pmatrix} \sim N\left[\begin{pmatrix} 0 \\ 0 \end{pmatrix}, \begin{pmatrix} 1 & 0.7 \\ 0.7 & 1 \end{pmatrix}\right]$$
$$X_{2,i} \sim N(0,1) \quad \text{for control units and} \quad X_{2,i} \sim N(0.25,1) \quad \text{for treated units},$$

where $X_1$ is also uniform across time, but it is correlated with a time-varying covariate, $X_5$, so we will introduce it later in this article. The correlations between covariates ($X_3$ and $X_4$, and $X_1$ and $X_5$) are calibrated to the correlations observed between covariates in the baseball example in Section 7 (i.e., height and weight have a correlation of approximately 0.7, and lag OBP and age have a correlation of approximately −0.4).

For treated units:

$$\begin{pmatrix} X_{1,i} \\ X_{5,i} \end{pmatrix} \sim N\left[\begin{pmatrix} 0 \\ 0 \end{pmatrix}, \begin{pmatrix} 1 & -0.4 \\ -0.4 & 1 \end{pmatrix}\right]$$
$$X_7, X_8 \sim N(0,1) \quad \text{and} \quad X_6 \sim N(0.5,1).$$

Four of the covariates are time-varying for control units. For each control unit, three instances are generated from a random walk process to correlate their values across time. Formally, for instance, $t$ in trajectory $i$, covariate $j$ is generated as follows:

$$\begin{pmatrix} X_{1,i} \\ X_{5,i,1} \end{pmatrix} \sim N\left[\begin{pmatrix} 0 \\ 0 \end{pmatrix}, \begin{pmatrix} 1 & -0.4 \\ -0.4 & 1 \end{pmatrix}\right]$$
$$X_{j,i,1} \sim N(0,1) \quad \text{for} \quad j = 6,7,8$$
$$X_{j,i,t} = X_{j,i,(t-1)} + \varepsilon_{j,i,(t-1)} \quad \text{for} \quad t = 2,3$$
$$\varepsilon_{j,i,1}, \varepsilon_{j,i,2} \sim N(0, 0.5^2).$$

Fixing $a_L = \log(1.25)$, $a_M = \log(2)$, $a_H = \log(4)$ and $a_{VH} = \log(10)$, and drawing the $\varepsilon_{i,t}$ terms independently from a standard normal distribution, we define outcomes as follows:

$$Y_{i,t} = a_{\mathrm{L}}\sum_{j=1}^{4} X_{j,i,t} + a_{VH}X_{5,i,t} + a_M(X_{6,i,t} + X_{8,i,t}) + a_H(X_{7,i,t}) + \Delta D_i + \varepsilon_{i,t}. \quad (2)$$

The outcome for a unit is correlated across time as it is generated from some time-varying covariates. Each simulation consists of 400 treated and 600 control individuals. We consider 1:2 matching. The true treatment effect, $\Delta$, is 0.25.

We consider two alternative ways of generating the continuous outcome variable besides model (2). First, we add correlation to the error terms within trajectories. Specifically, the $\varepsilon_{i,t}$s for a given trajectory $i$ are generated from a normal distribution with mean 0 and covariance matrix with off diagonal values of 0.8. Second, in addition to the correlated error terms, we square the $X_{2,i,t}$ term in the model, so it is no longer linear. We also run simulations with poor overlap. See Section C of the supplemental appendix for results and discussion of simulation performance under poor overlap.

We compare the bias-corrected block bootstrap approach outlined in Section 4, using a linear outcome model and a nonparametric bootstrap, to the confidence intervals obtained from WLS regression and WLS with clustered standard errors. We focus on the nonparametric bootstrap, as opposed to alternatives such as the wild bootstrap, because of its more common prevalence in practice; however, for comparisons between the wild bootstrap and the nonparametric bootstrap showing almost equivalent performance in a generally similar setting, see Otsu and Rai [15]. We choose to compare to WLS because this is commonly recommended in matching literature [33,34]. However, Abadie and Spiess [35] pointed out that standard errors from regression may be incorrect due to dependencies among outcomes of matched units, and identified matching with replacement as a setting in which these dependencies are particularly difficult to correct for. Our simulation results suggest that are used these difficulties carry over into the case of repeated measures. It is worth noting that the standard functions in R are used to compute WLS with matching weights such as lm and Zelig (which calls lm), compute biased standard error estimates in most settings. See Section B of the supplemental appendix for details.

## 5.2 Results

Tables 1 and 2 show the coverage and average 95% confidence interval (CI) length, respectively, of WLS regression, WLS regression with clustered standard errors, and bootstrap inference using our model-adjusted ATT estimator, for each of our three simulation settings under 10,000 simulations. As misspecification of the estimated linear outcome model increases, the bootstrap method is substantially more robust (although under substantial misspecification the bias-corrected method also fails to achieve nominal coverage). While the bootstrap confidence intervals are generally slightly wider than the WLS and WLS cluster confidence intervals, this is to be expected as the wider confidence intervals lead to improved coverage. In settings where strong scientific knowledge about the exact form of the outcome model is absent, the bootstrap approach appears more reliable than its chief competitors.

Results in Tables 1 and 2 are for GroupMatch with instance replacement, however matching with trajectory replacement performed very similarly in our simulations. Computation time was similar for GroupMatch with instance replacement and with trajectory replacement. In principle, GroupMatch with instance replacement should be substantially faster; however, in its current form GroupMatch does not implement the most computationally efficient algorithm for matching with instance replacement. Over 100 iterations, the average matching computation time was 4.63 seconds for matching with instance replacement and 4.71 seconds for matching with trajectory replacement. The average block bootstrap computation time was 2.51 seconds. Computation time was calculated on an Apple M1 Max 10-core CPU with 3.22 GHz processor and 64 GB RAM running on macOS Monterey.

**Table 1:** Coverage of the WLS, WLS cluster and bootstrap bias corrected methods of inference for our three simulation set-ups

| **Coverage** | **WLS** (%) | **WLS cluster** (%) | **Bootstrap bias corrected** (%) |
|---|---|---|---|
| Linear DGP | 92.6 | 94.4 | 94.0 |
| Linear DGP, correlated errors | 89.4 | 91.7 | 94.4 |
| Nonlinear DGP, correlated errors | 83.3 | 86.2 | 89.8 |

**Table 2:** Average 95% confidence interval length for the WLS, WLS cluster and bootstrap bias corrected methods of inference for our three simulation set-ups

| Average CI length | WLS | WLS cluster | Bootstrap bias corrected |
|---|---|---|---|
| Linear DGP | 0.25 | 0.27 | 0.27 |
| Linear DGP, correlated errors | 0.25 | 0.27 | 0.30 |
| Nonlinear DGP, correlated errors | 0.26 | 0.28 | 0.31 |

# 6 Testing for timepoint agnosticism

The key advantage of GroupMatch relative to other matching techniques designed for rolling enrollment settings is its ability to consider and optimize over matches between units at different timepoints, which leads to higher quality matches on lagged covariates. This advantage comes with a price in additional assumptions, notably the assumption of timepoint agnosticism. Timepoint agnosticism means that mean potential outcomes under control for any two individual timepoints in the data should be identical; in particular, this rules out time trends of any kind in the outcome model that cannot be explained by covariates in the prior $L$ timepoints.

While in many applications scientific intuition about the data generating process suggests this assumption may be reasonable, it is essential that we consider any information contained in the observed data about whether it holds in a particular case. Accordingly, we present a falsification test for timepoint agnosticism. Falsification tests are tests "for treatment effects in places where the analyst knows they should not exist" [36] and are useful in a variety of settings in observational studies [37]. In particular, our test is designed to detect violations of timepoint agnosticism, or "treatment effects of time" when they should be absent; rejections indicate settings in which GroupMatch is not advisable and other rolling enrollment matching techniques that do not rely on timepoint agnosticism are likely more suitable. While failure to reject may not constitute proof positive of timepoint agnosticism's validity, it rules out gross violations, thereby limiting the potential for bias.

To test the timepoint agnosticism assumption we use *control-control time matching*: matching control units at different timepoints and testing if the average difference in outcomes between the two timepoint groups, conditional on relevant covariates, is significantly different from zero using a bootstrap test. Specifically, restricting attention to trajectories $i$ from the control group, we select two timepoints $t_0$ and $t_1$ and match each instance at one timepoint to one at the other timepoint using the GroupMatch optimization routine, based on similarity of covariate histories over the previous $L$ timepoints. Since this match compares instances at two fixed time points, any optimal method of matching without replacement may be used. One practical issue arises: GroupMatch and related matching routines expect one group to be designated "treated," all members of which are generally retained in the match, and the other "control," some members of which will be included, but both matching groups are controls in this case. We label whichever of the two groups has fewer instances as treated; without loss of generality, we will assume there are fewer instances at time $t_1$ and use these instances as the reference group to be retained.

The test statistic for the falsification test is motivated by the ATT estimator in Section 2.2. Let $N_c$ be the total number of control units and let $N_{t_1}$ be the number of control instances at time $t_1$. Let $\hat{\mu}_0^{t_0}$ be a bias correction model fit on our new control group (i.e., control instances at time $t_0$). In addition, let $D_{it}' = 1$ if unit $i$ is present at time $t$. We define the test statistic as follows:

$$\begin{aligned}\hat{\Delta}_{cc} &= \frac{1}{N_{t_1}}\sum_{i=1}^{N_c} D'_{it_1}\left(\left(Y_{i,t=t_1} - \hat{\mu}_0^{t_0}\left(\mathbf{X}_{i,t=t_1}\right)\right) - \sum_{j=1}^{N_c} M_{it_1,jt_0}\left(Y_{j,t=t_0} - \hat{\mu}_0^{t_0}\left(\mathbf{X}_{j,t=t_0}\right)\right)\right)\\ &= \frac{1}{N_{t_1}}\sum_{i=1}^{N_c}\left[D'_{it_1}\left(Y_{i,t=t_1} - \hat{\mu}_0^{t_0}\left(\mathbf{X}_{i,t=t_1}\right)\right) - D'_{it_0}K_M(i,t_0)\left(Y_{i,t=t_0} - \hat{\mu}_0^{t_0}\left(\mathbf{X}_{i,t=t_0}\right)\right)\right]\\ &= \frac{1}{N_{t_1}}\sum_{i=1}^{N_c}\hat{\Delta}_{cc,i}.\end{aligned}$$

We use a bootstrap test to test the following null hypothesis, where $E_0^{t_1}\{\cdot\}$ indicates expectation over the distribution of the covariates in control instances at time $t_1$.

$$E_0^{t_1}\{\mu_0^{t_0}(\mathbf{X})\} = E_0^{t_1}\{\mu_0^{t_1}(\mathbf{X})\}.$$

In words, this null hypothesis says that, accounting for differences in the covariate distribution at times 0 and 1, the difference in the average outcomes of control instances at the two timepoints is zero.

The test constructs a bootstrap confidence interval as in Section 3 and checks whether the interval covers 0. If the interval covers 0, the test fails to reject. We present the following steps:

(1) Label control instances from the first group of trajectories at timepoint $t_1$ the new "treated" units, and control instances from the second group of trajectories at timepoint $t_0$ the new "control" units.
(2) Fit a bias correction model on the new control units.
(3) Match the new treated units to the new control units and calculate the test statistic.
(4) Repeat $B$ times:
    (a) Randomly resample $N_c$ elements $\hat{\Delta}*_{cc,i}$ with replacement from $\{\hat{\Delta}_{cc,1},\ldots,\hat{\Delta}_{cc,N_c}\}$.
    (b) Calculate $\hat{\Delta}_{cc}$ on the resampled data.
(5) Construct a $(1-\alpha)$ confidence interval based on the $\alpha/2$ and $1-\alpha/2$ percentile of the values calculated from the boostrap samples.
(6) If this confidence interval covers 0, fail to reject the null hypothesis.

We choose to use a bootstrap test here in line with our inference methods in previous sections. However, it is worth noting that a permutation test is also feasible here.

A key consideration for the falsification test is which timepoints to choose as $t_0$ and $t_1$. The choice of timepoint comparison depends largely on what a plausible time trend would be for the problem at hand. For example, if you suspect a linear time trend, it makes sense to look at the first and last timepoints. If the trend is linear, this test should have high power to detect a problem in moderate to large samples. If one is uncertain about the specific shape of the time trend that is most likely to occur and wants to test for all possible trends, we recommend testing each sequential pair of timepoints (i.e., timepoints 1 and 2, 2 and 3, 3 and 4, and so on) and using a multiplicity adjustment.

The falsification test is subject to several common criticisms levied at falsification tests, particularly their ineffectiveness in settings with low power. One possible approach is to reconfigure the test to assume violation of timepoint agnosticism as a null hypothesis and seek evidence in the data to reject it; Hartman and Hidalgo [38] recommend a similar change for falsification tests used to assess covariate balance, called *equivalence tests*.

The implementation of these modifications is fairly straightforward. First, we must define an equivalence range for our outcome variable: a set of values for which the difference is substantively inconsequential. Let $\varepsilon_L$ and $\varepsilon_U$ denote the lower and upper bounds under which the outcome variable is deemed equivalent. Hartman and Hidalgo [38] recommend using $\varepsilon = \pm 0.36\sigma$ as a default when researchers are unsure of an appropriate equivalence region. Next, before step 4 in our falsification test, one simply subtracts $\varepsilon_L$ (and in a separate run $\varepsilon_U$) from all treated outcomes. If either one-sided test fails to reject the null, then the test fails.

See Section D of the supplemental appendix for simulations illustrating this method.

# 7 Application: baseball injuries

We study the impact of short-term injury on hitting performance in observational data from MLB during 2013–2017. Quantitative studies of major league hitting performance [39] and of injury trends and impact in athletics [12] have been performed repeatedly, but only a few studies so far have evaluated the impact of injury on position players' hitting performance. These have focused on specific injury types and have not found strong evidence that injury is associated with a decline in performance [11,13,14].

We use GroupMatch to match baseball players injured at certain times to similar players at other points in the season that were not injured. We evaluate whether players see a decline in offensive performance immediately after their return from injury. In contrast to other studies, we pool across injury types to see if there is a more general effect of short-term injury on hitter performance.

## 7.1 Data and methodology

We use publicly available MLB player data from Retrosheet.org and injury data scraped from ProSportsTransactions.com for the years 2013–2017. Our dataset is composed of player height, weight and age, quantities that remain constant over a single season of play, as well as OBP, plate appearances (PAs) at different points in the season, and dates of short-term injuries, in which the player's team designated him for a 7–10 day stay on the team's official injured list, for each year. OBP is a common measure of hitter performance and is approximately equal to the number of times a player reaches base divided by their number of plate appearances.[1]

For each noninjured player, we generate three pseudo-injury dates evenly spaced over their PAs. In each season, we match injured players to four noninjured players. Matches were formed using GroupMatch with instance replacement, matching on age, weight, height, number of times previously injured, recent performance measured by OBP over the previous 100 PAs, and performance over the entire previous year as measured by end-of-year OBP after James-Stein shrinkage.[2] We choose to shrink OBP using James-Stein to limit the impact of sampling variability for players with a relatively small number of PAs the previous season [40]. Table 3 shows the balance for each of the covariates before and after matching. For each covariate, matching shrinks the standardized difference between the treated and control means. The balance achieved is not perfect, especially for the number of previous injuries. This underlines the importance of combining matching with bias correction to clean up imbalances not removed by matching.

## 7.2 Results

We compare the results for bias-corrected block bootstrap inference, WLS, and WLS with clustered standard errors. The ATT estimates are positive (0.010), but the 95% confidence intervals cover zero for all methods, indicating that there is not strong evidence that short-term injury impacts hitter performance. We present the results for 2017 in Figure 2. Results from each of 2013–2016 were substantively the same, as were results obtained by pooling the matched data across years. The data pass the timepoint agnosticism test, comparing the first and last pseudo-injury dates. We chose to compare the first and last pseudo-injury dates, because we were most concerned about player performance degrading over the course of the entire season due to fatigue. We also perform equivalence tests using $\varepsilon = \pm 0.02$, which our data also pass.

**1** OBP = (Hits + Walks + Hit By Pitch)/(At Bats + Walks + Hit by Pitch + Sacrifice Flies).

**2** See https://chris-said.io/2017/05/03/empirical-bayes-for-multiple-sample-sizes/for discussion of James-Stein shrinkage to estimators with variable sample sizes.

**Table 3:** Balance table for MLB injury analysis before and after matching each injured player to four noninjured players

| Variable | Treated mean | Control mean | | Standardized difference | |
|---|---|---|---|---|---|
| | | Before | After | Before | After |
| Height | 73.7 | 73.1 | 73.4 | 0.26 | 0.14 |
| Weight | 213 | 209 | 212 | 0.24 | 0.07 |
| 2016 OBP (JS shrunk) | 0.324 | 0.328 | 0.323 | −0.09 | 0.02 |
| Lag OBP | 0.336 | 0.341 | 0.338 | −0.07 | −0.02 |
| Birth year | 1988 | 1988 | 1988 | −0.08 | −0.06 |
| Number previous injuries | 2.73 | 1.91 | 2.16 | 0.30 | 0.21 |

We could have also chosen to use a difference-in-differences estimator here, using the difference in performance right before and after the injury, or pseudo-injury, date as our outcome. Results are substantively the same for both estimators. This similarity is due to the close lag OBP matches GroupMatch produces for this example.

## 8 Discussion

The introduction of GroupMatch with instance replacement, a method for block bootstrap inference, and a test for timepoint agnosticism provide substantial new capabilities for matching in settings with rolling enrollment. We now discuss a number of limitations and opportunities for improvement.

Our proof of the block bootstrap approach assumes the use of GroupMatch with instance replacement. The large-sample properties of matched-pair discrepancies are substantially easier to analyze mathematically in this setting than GroupMatch with trajectory replacement or GroupMatch without replacement, designs in which different treated units may compete for the same control units, and the technical argument must be altered to account for this complexity. However, Abadie and Imbens [27] successfully characterized similar large-sample properties in cross-sectional settings for matching without replacement. While beyond the scope of our work here, we believe it is likely that this approach could provide an avenue for extending Theorem 1 to cover the other two GroupMatch designs. Empirically, we have found that the block bootstrap performs well when matches are calculated using any of the three GroupMatch designs.

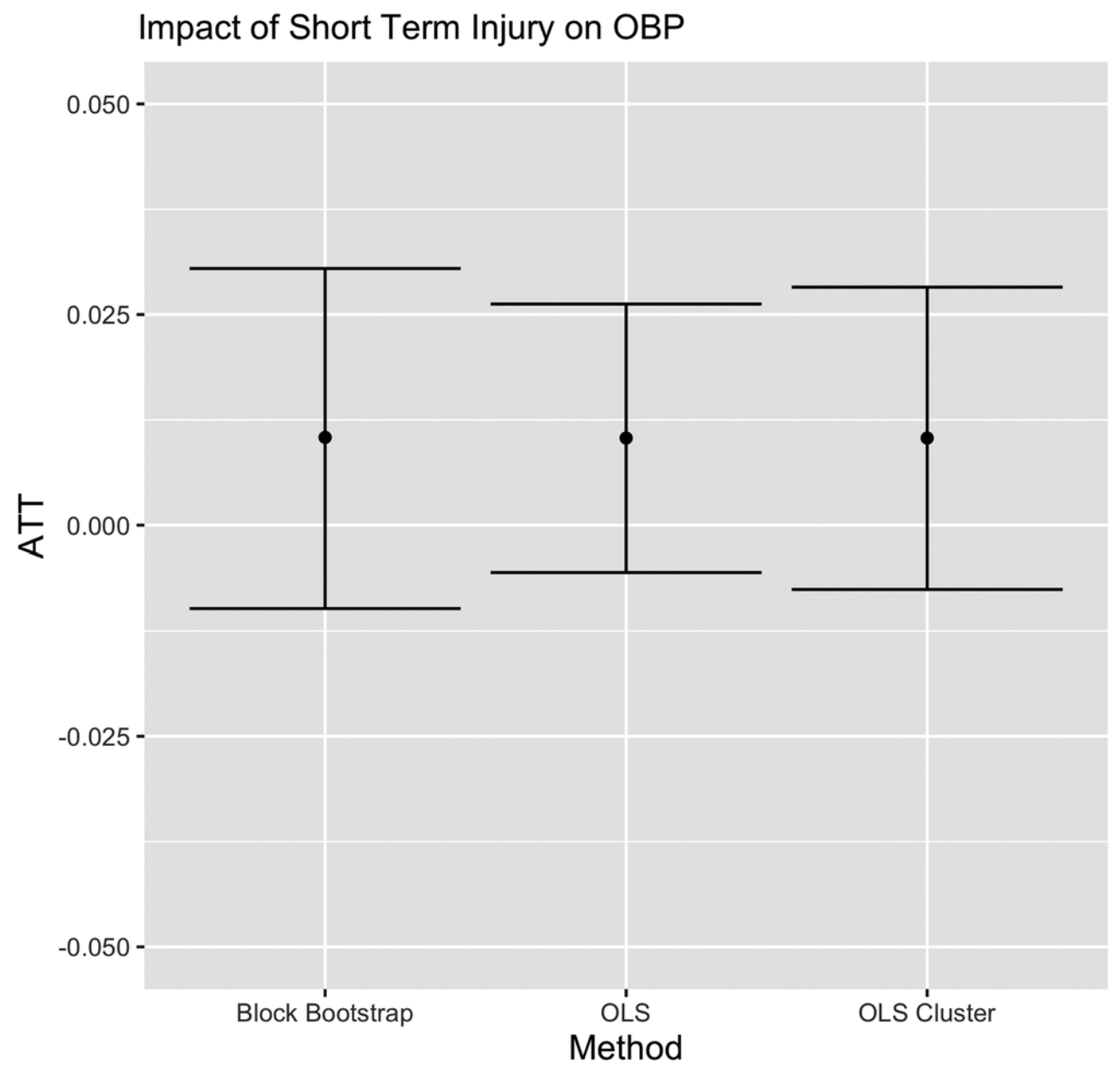

**Figure 2:** Estimates and 95% confidence intervals for block bootstrap, WLS and cluster WLS inference methods for the ATT in our 2017 baseball injury analysis.

Setting aside the technical barriers associated with extending the theory to GroupMatch without replacement, our new approach provides a competitor method to the existing randomization inference framework described by Pimentel et al. [10] available for GroupMatch without replacement. The randomization inference framework offers the advantage of closely-related methods of sensitivity analysis and freedom from making assumptions about the sampling distribution of the response variables; on the other hand, the block bootstrap method avoids the need to assume a sharp null hypothesis. In general, these same considerations arise in choosing between sampling-based inference and randomization-based inference for a cross-sectional matched study, although such choices have received surprisingly little direct and practical attention in the literature thus far.

As described in Section 6, the falsification test faces several criticisms, such as ineffectiveness in lower power settings. The modifications to construct equivalence tests as in Hartman and Hildago [38] address these concerns. However, even in the absence of such a change, the falsification test may prove useful in concert with a sensitivity analysis. Sensitivity analysis, already widely studied in causal inference as a way to assess the role of ignorability assumptions, places a nonzero bound on the degree of violation of an assumption and reinterprets the study's results under this bound, often repeating the process for larger and larger values of the bound to gain insight. Such a procedure, which focuses primarily on assessing the impact of small or bounded violations of an assumption, naturally complements our falsification test, which can successfully rule out large violations but is more equivocal about minor violations.

Unfortunately, no sensitivity analysis appropriate for block bootstrap inference has been developed yet, either for timepoint agnosticism or other strong assumptions such as ignorability. The many existing methods for sensitivity analysis (developed primarily with ignorability assumptions in mind) are unsatisfying in our framework for a variety of reasons: some rely on randomization inference [29], others focus on weighting methods rather than matching [41,42], and others are limited to specific outcome measures [43] or specific test statistics [44]. We view the development of compelling sensitivity analysis approaches to be an especially important methodological objective for matching under rolling enrollment.

Finally, we note that in cross-sectional settings, moderate imbalances like those observed after matching in the baseball study in Section 7 can often be removed by refining the match to include calipers [45,46] or balance constraints [47–49] on important variables. For computational reasons, these constraints are difficult to implement and use in full generality for GroupMatch designs. For example, some balance constraints rely on network flow representations of the matching problem that are not immediately compatible with the network flow representation underpinning GroupMatch. Further work to consider how calipers and balance constraints can be elegantly incorporated will enhance GroupMatch's effectiveness in practice.

**Acknowledgments:** The MLB performance data was obtained from and is copyrighted by Retrosheet (www.retrosheet.org). We thank the author and maintainer of the GitHub repository https://github.com/robotallie/baseball-injuriesfor making the injury data easily available. We also thank Eli Ben-Michael, Peng Ding, Avi Feller, Lauren Forrow, Shirshendu Ganguly, and Jiaqi Li for helpful conversations and feedback.

**Funding information**: Amanda Glazer (DMS RTG #1745640) and Samuel D. Pimentel (#2142146) acknowledge support from the National Science Foundation.

**Conflict of interest**: Prof. Samuel D. Pimentel is a member of the Editorial Board of the Journal of Causal Inference but was not involved in the review process of this article.

# Supplementary material

## A Proof of theorem

### A.1 Assumptions

Our proof relies on the assumptions on sampling described in Section 2.1 and the GroupMatch identification assumptions described in Section 2.2 of the main manuscript, with the exception of the exact matching assumption. In addition, we invoke three additional sets of assumptions, which we denote M, W, R following similar labeling in Otsu and Rai [1], from whom we adapt our proof strategy.

**Assumption M.** (**Conditions for $\hat{\Delta}_{\text{adj}}$**)

(1) Let the population distribution of $\mathbf{X}_i = (X_{i,1}, \ldots, X_{i,T})$ be continuous on $\mathbb{R}^{kT}$ with compact and convex support $\mathbb{X}^T$. In addition, let the density of $\mathbf{X}_i$ be bounded, and bounded away from zero on its support.

(2) For some $r \geq 1$, $\frac{N_1^r}{N_0} \to \theta$ for $\theta \in (0, \infty)$.

(3) For $z = 0, 1$, let $\mu_z^t(\mathbf{X})$ be Lipschitz in $\mathbb{X}^L$ for all $t = L + 1, \ldots, T$.

(4) For all $t \in \{L + 1, \ldots, T\}$, and $d = 0, 1$, $E[Y_{i,t}^4(z)|D_i = d, \mathbf{X}_i]$ and $\text{Cov}(Y_{it}, Y_{it'}|D_i = d, \mathbf{X}_{i,t} = x, \mathbf{X}_{i,t'} = x')$ are bounded uniformly on $\mathbb{X}^T$, and $\text{Var}(Y_{i,t}(z)\|D_i = d, \mathbf{X}_i)$ is Lipschitz in $\mathbb{X}^T$ and bounded away from zero.

**Assumption R.** (**Conditions for $\mu_d(x)$**)

For $d = 0, 1$ and $\lambda$ satisfying $\sum_{l=1}^{kL} \lambda_l = kL$, the derivative $\partial^{kL} \mu_d^t(x)$ exists and satisfies $\sup_{x \in \mathbb{X}} |\partial^{kL} \mu_d^t(x)| \leq R$ for some $R > 0$ and for all $t \in \{L + 1, \ldots T\}$. Furthermore, $\hat{\mu}_0(x)$ satisfies $|\hat{\mu}_0(\cdot) - \mu_0(\cdot)|_{kL-1} = o_p(N^{-1/2+1/(kL)})$.

**Assumption W.** (**Conditions for $W_i^*$**)

(1) $(W_1^*, \ldots, W_N^*)$ is exchangeable and independent of $\mathbf{O} = (\mathbf{Y,D,X})$.

(2) $\sum_{i=1}^N (W_i^* - \bar{W}^*)^2 \xrightarrow{p} 1$ where $\bar{W}^* = \frac{1}{N} \sum_{i=1}^N W_i^*$

(3) $\max_{i=1,\ldots,N} |W_i^* - \bar{W}^*| \xrightarrow{p} 0$

(4) $E[W_i^{*2}] = O(N^{-1})$ for all $i = 1, \ldots, N$

Assumption $W$ is satisfied by construction for the nonparametric bootstrap, wild bootstrap, and Bayesian bootstrap approaches mentioned in the main text.

### A.2 Lemmas

To bound the size of matching discrepancies, we compare those obtained by GroupMatch with instance replacement to nearest-neighbor matching at a fixed timepoint, in which only one instance from each control unit can potentially be used in a match. Nearest-neighbor matching matches each treated unit to the control instance that is most similar. While GroupMatch also uses nearest-neighbor matching, nearest-neighbor matching at a fixed timepoint considers a smaller pool of control instances since there is only one instance for each control unit that can potentially be used in a match. Comparing matching discrepancies between GroupMatch with instance replacement and nearest-neighbor matching is important to apply Lemma A.2, used to prove Theorem 2, in Abadie and Imbens [2]. For each treated unit $i$, let $j_m(i)$, and $j_m^{gm}(i)$ represent the $m$th instance used as a match for nearest-neighbors at a fixed timepoint and GroupMatch with instance replacement respectively. Let $U_{m,i} = \mathbf{X}_{j_m(i)} - \mathbf{X}_{i,T_i}$ and $U^{gm}_{m,i} = \mathbf{X}_{j^{gm}_m(i)} - \mathbf{X}_{i,T_i}$ be the matching discrepancies under nearest neighbors at a fixed timepoint and GroupMatch with instance replacement matching respectively.

**Lemma 1.** $\|U^{gm}_{m,i}\| \leq \|U_{m,i}\|$ *for all* $m, i$.

**Proof.** Because nearest neighbors (NN) matching at a fixed timepoint only considers one instance from each control trajectory whereas GroupMatch with instance replacement (GM) considers multiple, the set of control instances that can be used in a match in NN matching, $C$, is a subset of the set of control instances that can be used in a match in GM, $C^{gm}$: $C \subseteq C^{gm}$. Both NN and GM match treated units to the control instance that minimizes $U_{m,i}$ and $U^{gm}_{m,i}$, so we have that

$$\|U^{gm}_{m,i}\| = \min_{m, j \in C^{gm}} |\mathbf{X}_j - \mathbf{X}_{i,T_i}| \leq \min_{m, j \in C} |\mathbf{X}_j - \mathbf{X}_{i,T_i}| = \|U_{m,i}\|,$$

where $\min_m$ denotes the $m$th minimum. □

**Lemma 2.** $E[K_M(i,t)^q]$ *is bounded uniformly in N.*

**Proof.** This proof follows closely with the proof of Lemma 3 in Abadie and Imbens [3] with modifications described below. Let $f^t$ be the density of $\mathbf{X}_{i,t}$ and define $\underline{f} = \inf_{x,w,t} f^t_w(x)$ and $\overline{f} = \sup_{x,w,t} f^t_w(x)$. We define the catchment area, $\mathbb{A}_M(i,t)$ as the subset of $\mathbb{X}$ such that control unit $i$ at time $t$ is matched to each observation $j$ at time $t'$ with $W_j = 1 - W_i$ and $\mathbf{X}_{j,t'} \in \mathbb{A}_M(i,t)$:

$$\mathbb{A}_M(i,t) = \left\{ x \middle| \sum_{l | W_l = W_i, l \neq i} 1\{(\min_{t'} \|\mathbf{X}_{l,t'} - x\|) \leq \|\mathbf{X}_{i,t} - x\|\} 1\{\min_{t'} \|\mathbf{X}_{i,t'} - x\| \geq \|\mathbf{X}_{i,t} - x\|\} \leq M \right\}.$$

Ultimately, we want to bound the volume of the catchment area. In order to do this, we need to bound the probability that the distance to a match exceeds some value. To derive this bound, we must account for our trajectory structure by showing the following inequality:

$$\begin{aligned}
&\Pr\Big(\|\mathbf{X}_{j,t} - \mathbf{X}_{i,t'}\| > uN^{-1/k}_{1-W_i} | W_1, \ldots, W_N, \mathbf{X}_{i,t'} = x, j \in \mathcal{J}_M(i)\Big) \\
&\leq \Pr\Big(\|\mathbf{X}_{j,t} - \mathbf{X}_{i,t'}\| > uN^{-1/k}_{1-W_i} | W_1, \ldots, W_N, \mathbf{X}_{i,t'} = x, (j,t) = j_M(i,t')\Big) \\
&= \sum_{m=0}^{M-1} \binom{N_{1-W_i}}{m} \Pr\Big(\min_t \|\mathbf{X}_{j,t} - \mathbf{X}_{i,t'}\| > uN^{-1/k}_{1-W_i} | W_1, \ldots, W_N, W_j = 1 - W_i, \mathbf{X}_{i,t'} = x\Big)^{N_{1-W_i} - m} \\
&\quad \times \Pr\Big(\min_t \|\mathbf{X}_{j,t} - \mathbf{X}_{i,t'}\| \leq uN^{-1/k}_{1-W_i} | W_1, \ldots, W_N, W_j = 1 - W_i, \mathbf{X}_{i,t'} = x\Big)^m \\
&\leq \sum_{m=0}^{M-1} \binom{N_{1-W_i}}{m} \Pr\Big(\|\mathbf{X}_{j,t} - \mathbf{X}_{i,t'}\| > uN^{-1/k}_{1-W_i} | W_1, \ldots, W_N, W_j = 1 - W_i, \mathbf{X}_{i,t'} = x\Big)^{N_{1-W_i} - m} \\
&\quad \times \Pr\Big(\|\mathbf{X}_{j,t} - \mathbf{X}_{i,t'}\| \leq uN^{-1/k}_{1-W_i} | W_1, \ldots, W_N, W_j = 1 - W_i, \mathbf{X}_{i,t'} = x\Big)^m.
\end{aligned}$$

The second inequality follows from the fact that the probability that the minimal distance over a trajectory is less than or equal to any particular instance. The rest of the proof follows directly from the proof of Abadie and Imbens [3]'s Lemma 3 after substituting in our catchment area, this inequality, and indexing over time. □

**Lemma 3.** *Given that* $E[K_M(i,t)^q]$ *is uniformly bounded,* $\sum_i \sum_t K_M(i,t) = N_1$, *and* $K_M(i,t) \geq 0$, *we have that* $\sum_i \sum_t \sum_{t'} \sum_{t''} \sum_{t'''} E[K_M(i,t) K_M(i,t') K_M(i,t'') K_M(i,t''')] \leq c N_1^{1+o(1)}$ *for some constant* $c$.

**Proof.** For all $\varepsilon > 0$, we have that

$$\begin{aligned}
&\sum_i\sum_t\sum_{t'}\sum_{t''}\sum_{t'''}E[K_M(i,t)K_M(i,t')K_M(i,t'')K_M(i,t''')]\\
&= \sum_i\sum_t\sum_{t'}\sum_{t''}\sum_{t'''}E[K_M(i,t)K_M(i,t')K_M(i,t'')K_M(i,t''');K_M(i,t),K_M(i,t'),K_M(i,t''),K_M(i,t''')\le N_1^{\varepsilon}]\\
&\quad+\sum_i\sum_t\sum_{t'}\sum_{t''}\sum_{t'''}E\Big[K_M(i,t)K_M(i,t')K_M(i,t'')K_M(i,t''');\max_{t_1=t,t',t'',t'''}K_M(i,t_1)>N_1^{\varepsilon}\Big]\\
&\le \sum_i\sum_t\sum_{t'}\sum_{t''}\sum_{t'''}E[K_M(i,t)K_M(i,t')K_M(i,t'')K_M(i,t''');K_M(i,t),K_M(i,t'),K_M(i,t''),K_M(i,t''')\le N_1^{\varepsilon}]\\
&\quad+4\sum_i\sum_t\sum_{t'}\sum_{t''}\sum_{t'''}E[K_M(i,t)K_M(i,t')K_M(i,t'')K_M(i,t''');K_M(i,t_1)>N_1^{\varepsilon}]
\end{aligned}$$

This upper bound is derived from the fact that

$$\begin{aligned}
&E[K_M(i,t)K_M(i,t')K_M(i,t'')K_M(i,t''');\max_{t_1=t,t',t'',t'''}K_M(i,t_1)>N_1^{\varepsilon}]\\
&\le \sum_{t_1\in\{t,t',t'',t'''\}}E[K_M(i,t)K_M(i,t')K_M(i,t'')K_M(i,t''');K_M(i,t_1)>N_1^{\varepsilon}]
\end{aligned}$$

and since we are summing over all $t, t', t'', t'''$ we are able to replace the summation above with multiplying by four. We bound each of the terms separately. First,

$$\begin{aligned}
&\sum_i\sum_t\sum_{t'}\sum_{t''}\sum_{t'''}E[K_M(i,t)K_M(i,t')K_M(i,t'')K_M(i,t''');K_M(i,t),K_M(i,t'),K_M(i,t''),K_M(i,t''')\le N_1^{\varepsilon}]\\
&\le N_1^{3\varepsilon}\sum_i\sum_t\sum_{t'}\sum_{t''}\sum_{t'''}E[K_M(i,t)] = T^3N_1^{3\varepsilon+1}
\end{aligned}$$

For the next term we use proof by contradiction to show

$$\sum_i\sum_t\sum_{t'}\sum_{t''}\sum_{t'''}E[K_M(i,t)K_M(i,t')K_M(i,t'')K_M(i,t''');\ K_M(i,t)>N_1^{\varepsilon}]\le N_1.$$

Suppose $\sum_i\sum_t\sum_{t'}\sum_{t''}\sum_{t'''}E[K_M(i,t)K_M(i,t')K_M(i,t'')K_M(i,t''');K_M(i,t)>N_1^{\varepsilon}]>N_1$, then:

$$\begin{aligned}
&\sum_i\sum_t\sum_{t'}\sum_{t''}\sum_{t'''}E[K_M(i,t)^{\frac{r}{\varepsilon}+1}K_M(i,t')K_M(i,t'')K_M(i,t''');K_M(i,t)>N_1^{\varepsilon}]\\
&\ge (N_1^{\varepsilon})^{\frac{r}{\varepsilon}}\sum_i\sum_t\sum_{t'}\sum_{t''}\sum_{t'''}E[K_M(i,t)K_M(i,t')K_M(i,t'')K_M(i,t''');K_M(i,t)>N_1^{\varepsilon}]\\
&= N_1^{r}\sum_i\sum_t\sum_{t'}\sum_{t''}\sum_{t'''}E[K_M(i,t)K_M(i,t')K_M(i,t'')K_M(i,t''');K_M(i,t)>N_1^{\varepsilon}]\\
&> N_1^rN_1 = N_1^{r+1}\\
&\Rightarrow NT^4\sup_{i,t,t',t'',t'''}E[K_M(i,t)^{\frac{r}{\varepsilon}+1}K_M(i,t')K_M(i,t'')K_M(i,t''');K_M(i,t)>N_1^{\varepsilon}]>N_1^{r+1}\\
&\Rightarrow \sup_{i,t,t',t'',t'''}E[K_M(i,t)^{\frac{r}{\varepsilon}+1}K_M(i,t')K_M(i,t'')K_M(i,t''');K_M(i,t)>N_1^{\varepsilon}]>\frac{N_1^{r+1}}{NT^4}=cN_1
\end{aligned}$$

for some constant $c$. This follows from Assumption $M$.

So, then we have that:

$$\begin{aligned}
\sup_{i,t}E[K_M(i,t)^{\frac{r}{\varepsilon}+4}] &\ge \sup_{i,t,t',t'',t'''}E[K_M(i,t)^{\frac{r}{\varepsilon}+1}K_M(i,t')K_M(i,t'')K_M(i,t''')]\\
&> \sup_{i,t,t',t'',t'''}E\Big[K_M(i,t)^{\frac{r}{\varepsilon}+1}K_M(i,t')K_M(i,t'')K_M(i,t''');K_M(i,t)>N_1^{\varepsilon}\Big]>cN_1
\end{aligned}$$

The first inequality holds by applying Cauchy-Schwarz twice. But then $E[K_M(i,t)^q]$ is not uniformly bounded for all $q$, so by contradiction

$$\sum_i\sum_t\sum_{t'}\sum_{t''}\sum_{t'''}E[K_M(i,t)K_M(i,t')K_M(i,t'')K_M(i,t''');K_M(i,t)>N_1^{\varepsilon}]\le N_1.$$

So we have that:

$$\begin{aligned}
&\sum_i\sum_t\sum_{t'}\sum_{t''}\sum_{t'''}E[K_M(i,t)K_M(i,t')K_M(i,t'')K_M(i,t''')] \\
&= \sum_i\sum_t\sum_{t'}\sum_{t''}\sum_{t'''}E[K_M(i,t)K_M(i,t')K_M(i,t'')K_M(i,t''');\ K_M(i,t),K_M(i,t'),K_M(i,t''),K_M(i,t''')\le N_1^{\varepsilon}] \\
&\quad+ \sum_i\sum_t\sum_{t'}\sum_{t''}\sum_{t'''}E[K_M(i,t)K_M(i,t')K_M(i,t'')K_M(i,t''');\ \max_{t_1=t,t',t'',t'''}K_M(i,t_1)>N_1^{\varepsilon}] \\
&\le \sum_i\sum_t\sum_{t'}\sum_{t''}\sum_{t'''}E[K_M(i,t)K_M(i,t')K_M(i,t'')K_M(i,t''');\ K_M(i,t),K_M(i,t'),K_M(i,t''),K_M(i,t''')\le N_1^{\varepsilon}] \\
&\quad+ 4\sum_i\sum_t\sum_{t'}\sum_{t''}\sum_{t'''}E[K_M(i,t)K_M(i,t')K_M(i,t'')K_M(i,t''');K_M(i,t)>N_1^{\varepsilon}] \\
&\le T^3N_1^{3\varepsilon+1}+4c'N_1\le cN_1^{1+o(1)}
\end{aligned}$$

for some constant $c$. □

## A.3 Proof

We follow the proof of Theorem 1 in Otsu and Rai [1] closely, adapting where necessary to address the possible presence of multiple correlated potential outcomes from the same trajectory in the matched design. In the case where every control unit has only one instance, our argument reduces to exactly that presented in Otsu and Rai [1].

We decompose $\sqrt{N_1}\,U^*$ as follows:

$$\begin{aligned}
\sqrt{N_1}\,U^* &= \sum_{i=1}^{N}W_i^*(\hat{\Delta}_i - D_i\hat{\Delta}_{\text{adj}}) \\
&= \sum_{i=1}^{N}(W_i^* - \bar{W}^*)(\hat{\Delta}_i - D_i\hat{\Delta}_{\text{adj}}) \\
&= \sum_{i=1}^{N}(W_i^* - \bar{W}^*)(D_i(\hat{\Delta}_i - \hat{\Delta}_{\text{adj}}) + (1-D_i)\hat{\Delta}_i) \\
&= \sum_{i=1}^{N}(W_i^* - \bar{W}^*)\left[D_i\left(Y_{i,T_i} - \hat{\mu}_0\left(\mathbf{X}_{i,T_i}\right) - \hat{\Delta}_{\text{adj}}\right) + (1-D_i)\sum_{t=1}^{T}\frac{K_M(i,t)}{C}(Y_{i,t} - \hat{\mu}_0(\mathbf{X}_{i,t}))\right] \\
&= \sqrt{N_1}\left(T^* + R_{1N_1}^* + R_{2N_1}^*\right)
\end{aligned}$$

We define the following:

$$\begin{aligned}
e_{i,t} &= Y_{i,t} - \mu_{D_i}(\mathbf{X}_{i,t}) \\
\xi_{i,t} &= (2D_i - 1)\left(\mu_{D_i}(\mathbf{X}_{i,t}) - \mu_{1-D_i}(\mathbf{X}_{i,t})\right) - \Delta
\end{aligned}$$

We can now rewrite the three components as follows.

$$\begin{aligned}
\sqrt{N_1}\,T^* &= \sum_{i=1}^{N}(W_i^* - \bar{W}^*)\left(D_i\left(e_{i,t=T_i} + \xi_{i,t=T_i}\right) - (1-D_i)\sum_{t=1}^{T}\frac{K_M(i,t)}{C}e_{i,t}\right) \\
&= \sum_{i=1}^{N}(W_i^* - \bar{W}^*)\left[D_i\left(\left(Y_{i,t=T_i} - \mu_1^{T_i}(\mathbf{X}_{i,t})\right) + \left(\mu_1^{T_i}(\mathbf{X}_{i,t}) - \mu_0(\mathbf{X}_{i,t}) - \Delta\right)\right) - (1-D_i)\sum_{t=1}^{T}\frac{K_M(i,t)}{C}(Y_{i,t} - \mu_0(\mathbf{X}_{i,t}))\right] \\
&= \sum_{i=1}^{N}(W_i^* - \bar{W}^*)\left[D_i\left(Y_{i,t=T_i} - \mu_0(\mathbf{X}_{i,t}) - \Delta\right) - (1-D_i)\sum_{t=1}^{T}\frac{K_M(i,t)}{C}(Y_{i,t} - \mu_0(\mathbf{X}_{i,t}))\right]
\end{aligned}$$

$$\sqrt{N_1}R^*_{1N_1} = \sum_{i=1}^{N}(W_i^* - \bar{W}^*)\left(D_i(\mu_0(\mathbf{X}_{i,t}) - \hat{\mu}_0(\mathbf{X}_{i,t})) - (1 - D_i)\frac{K_M(i,t)}{C}\sum_{t=1}^{T}(\mu_0(\mathbf{X}_{i,t}) - \hat{\mu}_0(\mathbf{X}_{i,t}))\right)$$

$$\sqrt{N_1}R^*_{2N_1} = \sum_{i=1}^{N}(W_i^* - \bar{W}^*)D_i(\Delta - \hat{\Delta}_{\text{adj}})$$

We have that $\Pr\{\sqrt{N_1}R^*_{1N_1} > \varepsilon|\mathbf{O}\} \xrightarrow{p} 0$ and $\Pr\{\sqrt{N_1}R^*_{2N_1} > \varepsilon|\mathbf{O}\} \xrightarrow{p} 0$ for any $\varepsilon > 0$, by the same argument as Otsu and Rai [1] which utilizes our Assumptions W, R, Lemma 2 and the Markov Inequality. This part of the argument also relies on Lemma 1, although the reliance is not explicit in Otsu and Rai [1]; see Abadie and Imbens [2] for a similar derivation where the role of Lemma 1 is more clear.

Next, to show that that $\sup_r|\Pr\{\sqrt{N_1}T^* \le r|\mathbf{O}\} - \Pr\{\sqrt{N_1}(\hat{\Delta}_{\text{adj}} - \Delta) \le r\}| \xrightarrow{p} 0$, we define:

$$\eta_i = \left[D_i\left(Y_{i,t=T_i} - \mu_0(\mathbf{X}_{i,t}) - \Delta\right) - (1 - D_i)\sum_{t=1}^{T}\frac{K_M(i,t)}{C}(Y_{i,t} - \mu_0(\mathbf{X}_{i,t}))\right] \Big/ \sqrt{N_1}$$
$$= D_i\left(e_{i,t=T_i} + \xi_{i,t=T_i}\right) - (1 - D_i)\sum_{t=1}^{T}\frac{K_M(i,t)}{C}e_{i,t}$$

We have the following:

$$\sigma_N^2 = \sigma_{1N}^2 + \sigma_2^2$$
$$\sigma_{1N}^2 = \frac{1}{N_1}\sum_{i=1}^{N}\mathrm{V\,ar}\left(\sum_{t=1}^{T}\left(D_i 1\{t = T_i\} + \frac{K_M(i,t)}{C}(1 - D_i)\right)Y_{it}|\mathbf{D}, \mathbf{X}\right)$$
$$= \frac{1}{N_1}\sum_{i=1}^{N}\mathrm{Cov}\left\{\sum_{t=1}^{T}(D_i 1\{t = T_i\} + \frac{K_M(i,t)}{C}(1 - D_i))Y_{it}, \sum_{t=1}^{T}(D_i 1\{t = T_i\} + \frac{K_M(i,t)}{C}(1 - D_i))Y_{it}|\mathbf{D}, \mathbf{X}\right\}$$
$$= \frac{1}{N_1}\sum_{i=1}^{N}\sum_{t=1}^{T}\sum_{t'=1}^{T}\left(D_i 1\{t = T_i\} + \frac{K_M(i,t)}{C}(1 - D_i)\right)\left(D_i 1\{t' = T_i\} + \frac{K_M(i,t')}{C}(1 - D_i)\right)\mathrm{Cov}(Y_{it}, Y_{it'})$$
$$\sigma_2^2 = E\left[\left(\left(\mu_1^{T_i}(\mathbf{X}_{i,t}) - \mu_0(\mathbf{X}_{i,t})\right) - \Delta\right)^2|D_i = 1\right]$$

Within $\sigma_{1N}^2$, note that $\mathrm{Var}(\sum_{t=1}^{T}(D_i 1\{t = T_i\} + \frac{K_M(i,t)}{C}(1 - D_i))Y_{it}|\mathbf{D}, \mathbf{X})$ reduces to $\mathrm{V\,ar}(Y_{it})$ for treated units. However, for control units, we have this variance term plus extra covariance terms between that control unit instance and the other instances in its trajectory. From here, we are able to follow the same proof strategy as in Otsu and Rai [1], with our modified assumptions, and some modifications (detailed below) to Lemmas (i)–(iii) in Otsu and Rai [1] to account for the extra covariance terms resulting from the trajectory structure of the control units. Lemmas (i)–(iii) in Otsu and Rai [1] show that the sampling variance of the $\eta_i$'s converge to the population variance, $\sigma_N^2$, and that other random variables converge in probability to 0. They show this by leveraging the boundedness of higher order moments of $\eta_i$ and use of the Markov inequality. We are able to utilize the same arguments, with changes laid out more explicitly for lemma (i), below, and generalized for lemmas (ii) and (iii).

To apply Lemmas (i)–(iii) in Otsu and Rai [1] we define:

$$\hat{\sigma}_{1N}^2 = \frac{1}{N_1}\sum_{i=1}^{N}\sum_{t=1}^{T}\sum_{t'=1}^{T}\left(D_i 1\{t = T_i\} + \frac{K_M(i,t)}{C}(1 - D_i)\right)\left(D_i 1\{t' = T_i\} + \frac{K_M(i,t')}{C}(1 - D_i)\right)e_{i,t}e_{i,t'}$$

Then note that:

$$
\begin{aligned}
E[(\hat{\sigma}_{1N}^2 - \sigma_{1N}^2)^2] = E\Bigg\{ & \Bigg( \frac{1}{N_1} \sum_{i=1}^{N} \sum_{t=1}^{T} \sum_{t'=1}^{T} (D_i 1\{t = T_i\} + \frac{K_M(i,t)}{C}(1 - D_i))(D_i 1\{t' = T_i\} + \frac{K_M(i,t')}{C}(1 - D_i)) e_{i,t} e_{i,t'} \Bigg) \\
& - \Bigg( \frac{1}{N_1} \sum_{i=1}^{N} \sum_{t=1}^{T} \sum_{t'=1}^{T} (D_i 1\{t = T_i\} + \frac{K_M(i,t)}{C}(1 - D_i))(D_i 1\{t' = T_i\} \\
& + \frac{K_M(i,t')}{C}(1 - D_i)) \mathrm{Cov}(Y_{it}, Y_{it'}) \Bigg) \Bigg\}^2 \\
= \frac{1}{N_1^2} E\Bigg[ & \Bigg\{ \sum_{i=1}^{N} \sum_{t=1}^{T} \sum_{t'=1}^{T} (D_i 1\{t = T_i\} + \frac{K_M(i,t)}{C}(1 - D_i)) \Big( D_i 1\{t' = T_i\} + \frac{K_M(i,t')}{C}(1 - D_i) \Big) \times (e_{i,t} e_{i,t'} - \mathrm{Cov}(Y_{it}, Y_{it'})) \Bigg\}^2 \Bigg] \\
= \frac{1}{N_1^2} E\Bigg[ & \sum_{i=1}^{N} \sum_{t=1}^{T} \sum_{t'=1}^{T} \sum_{j=1}^{N} \sum_{t_1=1}^{T} \sum_{t_1'=1}^{T} \Big( D_i 1\{t = T_i\} + \frac{K_M(i,t)}{C}(1 - D_i) \Big) \Big( D_i 1\{t' = T_i\} + \frac{K_M(i,t')}{C}(1 - D_i) \Big) \\
& \times \Big( D_j 1\{t_1 = T_j\} + \frac{K_M(j,t_1)}{C}(1 - D_j) \Big) \Big( D_j 1\{t_1' = T_j\} + \frac{K_M(j,t_1')}{C}(1 - D_j) \Big) \\
& \times (e_{i,t} e_{i,t'} - \mathrm{Cov}(Y_{it}, Y_{it'})) \Big( e_{j,t_1} e_{j,t_1'} - \mathrm{Cov}\big(Y_{jt_1}, Y_{jt_1'}\big) \Big) \Bigg] \\
= \frac{1}{N_1^2} & \sum_{i=1}^{N} \sum_{t=1}^{T} \sum_{t'=1}^{T} \sum_{t_1=1}^{T} \sum_{t_1'=1}^{T} E\Bigg[ \Big( D_i 1\{t = T_i\} + \frac{K_M(i,t)}{C}(1 - D_i) \Big) \Big( D_i 1\{t' = T_i\} + \frac{K_M(i,t')}{C}(1 - D_i) \Big) \\
& \times \Big( D_i 1\{t_1 = T_i\} + \frac{K_M(i,t_1)}{C}(1 - D_i) \Big) \Big( D_i 1\{t_1' = T_i\} + \frac{K_M(i,t_1')}{C}(1 - D_i) \Big) \\
& \times E[(e_{i,t} e_{i,t'} - \mathrm{Cov}(Y_{it}, Y_{it'})) \Big( e_{i,t_1} e_{i,t_1'} - \mathrm{Cov}\big(Y_{it_1}, Y_{it_1'}\big) \Big) | D_i, \mathbf{X}_{i,t}, \mathbf{X}_{i,t'}, \mathbf{X}_{i,t_1}, \mathbf{X}_{i,t_1'}] \Bigg] \\
\le \frac{1}{N_1^2} & (N_1 + c N_1^{1+o(1)}) \times 3 \times \max(\sup_{d,x} E[e_{it}^4 | D_i = d, \mathbf{X}_{i,t} = x], -2 \sup_{d,x,x'} E[e_{it}^2 | D_i = d, \mathbf{X}_{i,t} \\
& = x] \sup_{d,x,x'} \mathrm{Cov}(Y_{it}, Y_{it'} | D_i = d, \mathbf{X}_{i,t} = x, \mathbf{X}_{i,t'} = x'), \sup_{d,x,x'} \mathrm{Cov}(Y_{it}, Y_{it'} | D_i = d, \mathbf{X}_{i,t} = x, \mathbf{X}_{i,t'} = x')^2) \to 0.
\end{aligned}
$$

The inequality follows from Lemma 3. The convergence follows from Assumption M(4) and Lemma 2.

### A.3.1 Difference-in-differences estimator

This proof extends easily to the difference-in-differences ATT estimator described in Section 4.3 of the main manuscript with the modifications described below.

We replace $\hat{\Delta}_{\text{adj}}$ with $\hat{\Delta}_{DiD}$ and the outcome $Y_{i,t}$ with the difference-in-differences outcome $Y_{i,t} - Y_{i,t-1}$ We modify

$$e_{i,t} = Y_{i,t} - Y_{i,t-1} - \big( \mu_{D_i}(\mathbf{X}_{i,t}) - \mu_{D_i}(\mathbf{X}_{i,t-1}) \big)$$

and

$$\xi_{i,t} = (2D_i - 1)\Big( \big( \mu_{D_i}(\mathbf{X}_{i,t}) - \mu_{1-D_i}(\mathbf{X}_{i,t}) \big) - \big( \mu_{D_i}(\mathbf{X}_{i,t-1}) - \mu_{1-D_i}(\mathbf{X}_{i,t-1}) \big) \Big)$$

.

We also update our variance formulas to reflect the additional covariance terms introduced by the difference-in-differences outcome. In particular, our formula for $\sigma_{1N}^2$ includes additional terms:

$$\begin{aligned}
\sigma_{1N}^2 &= \frac{1}{N_1}\sum_{i=1}^{N} \mathrm{V\,ar}\left(\sum_{t=1}^{T}\left(D_i 1\{t = T_i\} + \frac{K_M(i,t)}{C}(1 - D_i)\right)(Y_{it} - Y_{i,t-1})|\mathbf{D},\mathbf{X}\right) \\
&= \frac{1}{N_1}\sum_{i=1}^{N} \mathrm{Cov}\left\{\sum_{t=1}^{T}\left(D_i 1\{t = T_i\} + \frac{K_M(i,t)}{C}(1 - D_i)\right)(Y_{it} - Y_{i,t-1}), \sum_{t=1}^{T}(D_i 1\{t = T_i\}\right. \\
&\quad \left. + \frac{K_M(i,t)}{C}(1 - D_i)\right)(Y_{it} - Y_{i,t-1})|\mathbf{D},\mathbf{X}\Big\} \\
&= \frac{1}{N_1}\sum_{i=1}^{N}\sum_{t=1}^{T}\sum_{t'=1}^{T}(D_i 1\{t = T_i\} + \frac{K_M(i,t)}{C}(1 - D_i))(D_i 1\{t' = T_i\} + \frac{K_M(i,t')}{C}(1 - D_i)) \\
&\quad \times (\mathrm{Cov}(Y_{it}, Y_{it'}) - \mathrm{Cov}(Y_{it}, Y_{i,t'-1}) - \mathrm{Cov}(Y_{i,t-1}, Y_{it'}) + \mathrm{Cov}(Y_{i,t-1}, Y_{i,t'-1}))
\end{aligned}$$

To show that $E[(\hat{\sigma}_{1N}^2 - \sigma_{1N}^2)^2] \to 0$ we use a similar argument to before, where we are able to bound the difference between the $e_{i,t}e_{i,t'}$ and $\mathrm{Cov}(Y_{it}, Y_{it'})$ by a fourth order polynomial in $e_{it}$ terms and use the assumption that the expectation of these fourth moments is bounded.

# B Weighted least squares

Weighted least squares (WLS) is commonly used with matching weights to calculate the ATT and its corresponding confidence interval [4,5]. For WLS, we have the following estimator for $\beta$:

$$\hat{\beta} = (X^T W X)^{-1} X^T W Y$$

$W$ is a diagonal matrix with entries corresponding to the matching weights. Note that $W = W^T$. We can compute the variance as follows:

$$\mathrm{V\,ar}(\hat{\beta}) = (X^T W X)^{-1} X^T W [\mathrm{V\,ar}(Y)]((X^T W X)^{-1} X^T W)^T$$

Software to compute WLS estimates, such as lm and Zelig (which calls on lm) in R, often assumes that $\mathrm{V\,ar}(Y) = \sigma^2 W^{-1}$. If this is true we get a cancellation that yields a nicer formula for variance:

$$\begin{aligned}
\mathrm{V\,ar}(\hat{\beta}) &= (X^T W X)^{-1} X^T W \sigma^2 W^{-1}((X^T W X)^{-1} X^T W)^T \\
&= (X^T W X)^{-1} X^T W W^{-1} W^T X \sigma^2 (W^T W X)^{-1} \\
&= (X^T W X)^{-1} (X^T W X) \sigma^2 (W^T W X)^{-1} \\
&= \sigma^2 (W^T W X)^{-1}
\end{aligned}$$

In R, lm (and Zelig) use this formula[1] in order to compute standard errors and confidence intervals for WLS regression. However, when this assumption is not true, as is the case in our simulations (and often in practice), this formula is incorrect.

Assume $\mathrm{V\,ar}(Y) = \sigma^2 I$, as in some of our simulations, then:

$$\mathrm{V\,ar}(\hat{\beta}) = (X^T W X)^{-1} X^T W \sigma^2 I((X^T W X)^{-1} X^T W)^T = \sigma^2 (X^T W X)^{-1} X^T W^2 X (X^T W X)^{-1}$$

To get an unbiased estimator of $\sigma^2$ we note that:

$$E[e^T I e] = \mathrm{tr}\left(I \textstyle\sum_{ee}\right)$$

Where $e = Y - \hat{Y} = Y - X\hat{\beta} = Y - X(X^T W X)^{-1} X^T W Y$. Now,

1 Last verified 04-11-2022.

**Table A1:** Average standardized difference between matched treatment and control groups for $X_2$ and $X_6$ for our simulation set-up under the bad overlap specification

| | $X_2$ | $X_6$ |
|---|---|---|
| Average Difference | 0.52 | 0.93 |

**Table A2:** Coverage of the WLS, WLS cluster and bootstrap bias corrected methods of inference for our three simulation set-ups under our bad overlap specification

| Coverage | WLS (%) | WLS cluster (%) | Bootstrap bias corrected (%) |
|---|---|---|---|
| Linear DGP, correlated errors | 77.3 | 87.3 | 73.7 |
| Nonlinear DGP, correlated errors | 71.1 | 83.1 | 73.7 |

**Table A3:** Average 95% confidence interval length for the WLS, WLS cluster and bootstrap bias corrected methods of inference for our three simulation set-ups under our bad overlap specification

| Average CI Length | WLS | WLS cluster | Bootstrap bias corrected |
|---|---|---|---|
| Linear DGP | 0.30 | 0.40 | 0.35 |
| Linear DGP, Correlated Errors | 0.31 | 0.39 | 0.32 |
| Nonlinear DGP, Correlated Errors | 0.32 | 0.41 | 0.44 |

$$\begin{aligned}\sum\nolimits_{ee} &= \mathrm{Cov}(Y - X(X^TWX)^{-1}X^TWY,\ Y - X(X^TWX)^{-1}X^TWY)\\ &= \mathrm{Var}(Y) - 2\mathrm{Cov}(Y, X(X^TWX)^{-1}X^TWY) + \mathrm{V\ ar}(X(X^TWX)^{-1}X^TWY)\\ &= \sigma^2 I - 2X(X^TWX)^{-1}X^TW\mathrm{V\ ar}(Y) + (X(X^TWX)^{-1}X^TW)^T(X(X^TWX)^{-1}X^TW)\mathrm{V\ ar}(Y)\\ &= \sigma^2(I - 2X(X^TWX)^{-1}X^TW + WX(X^TWX)^{-1}X^TX(X^TWX)^{-1}X^TW)\end{aligned}$$

Thus, to get an unbiased estimator of $\sigma^2$ we must divide $e^Te$ by the trace of $I - 2X(X^TWX)^{-1}X^TW + WX(X^TWX)^{-1}X^TX(X^TWX)^{-1}X^TW$. When we use this formula to create confidence intervals for WLS, instead of using lm, our simulations cover in the linear DGP, uncorrelated errors case. However, we do not expect that this correction addresses all issues with model-based parametric standard errors after matching under rolling enrollment — in particular, many of the problems identified by Abadie and Spiess [6] likely persist in some form — and we recommend use of the block bootstrap procedure discussed in the main manuscript instead.

**Table A4:** Summary of timepoint agnosticism simulation results

| Time trend, $\gamma$ | 0 | 0.1 | 0.25 |
|---|---|---|---|
| Proportion $P$-values <0.05 | 0.05 | 0.60 | 1 |

Proportion of 1,000 simulations where the $P$-value from the test is less than 0.05, for time trends, $\gamma = 0, 0.1, 0.25$.

# C Additional simulations

In this section, we provide additional simulations with the same set-up as Section 5 but worse overlap between treatment and control units.

In particular, we modify the distribution of $X_2$ and $X_6$ by increasing the mean for treated units by a multiple of 4. Now, for treated units:

$$X_2 \sim N(1, 1) \quad \text{and} \quad X_6 \sim N(2, 1)$$

The rest of the simulation set-up remains the same. Table A1 summarizes the average standardized difference in $X_2$ and $X_6$ between matched treatment and control groups. We do not include the other covariates in this table, because they are well matched with average differences close to 0. As evidenced by the table, this simulation set up has poor overlap leading to bad balance after matching.

Tables A2 and A3 summarize the coverage and average confidence interval length results for these simulations. Unsurprisingly, WLS Cluster continues to perform well in the linear DGP settings, even with bad overlap, because it has the correct model form. However, WLS and WLS Cluster’s coverage decreases substantially with increased misspecification. On the other hand, our bootstrap method does not perform as well because of the low quality of the matches. In general, we’d advise against trusting matching results when there is high imbalance in the matched data.

# D Falsification test simulations

In this section, we illustrate the timepoint agnosticism falsification test presented in Section 6 of the main minuscript via simulation. We generate a dataset of 1,000 control units each with 4 covariates and 2 instances occuring at times $t_0$ and $t_1$. Two of the covariates vary with time, and two are uniform across time:

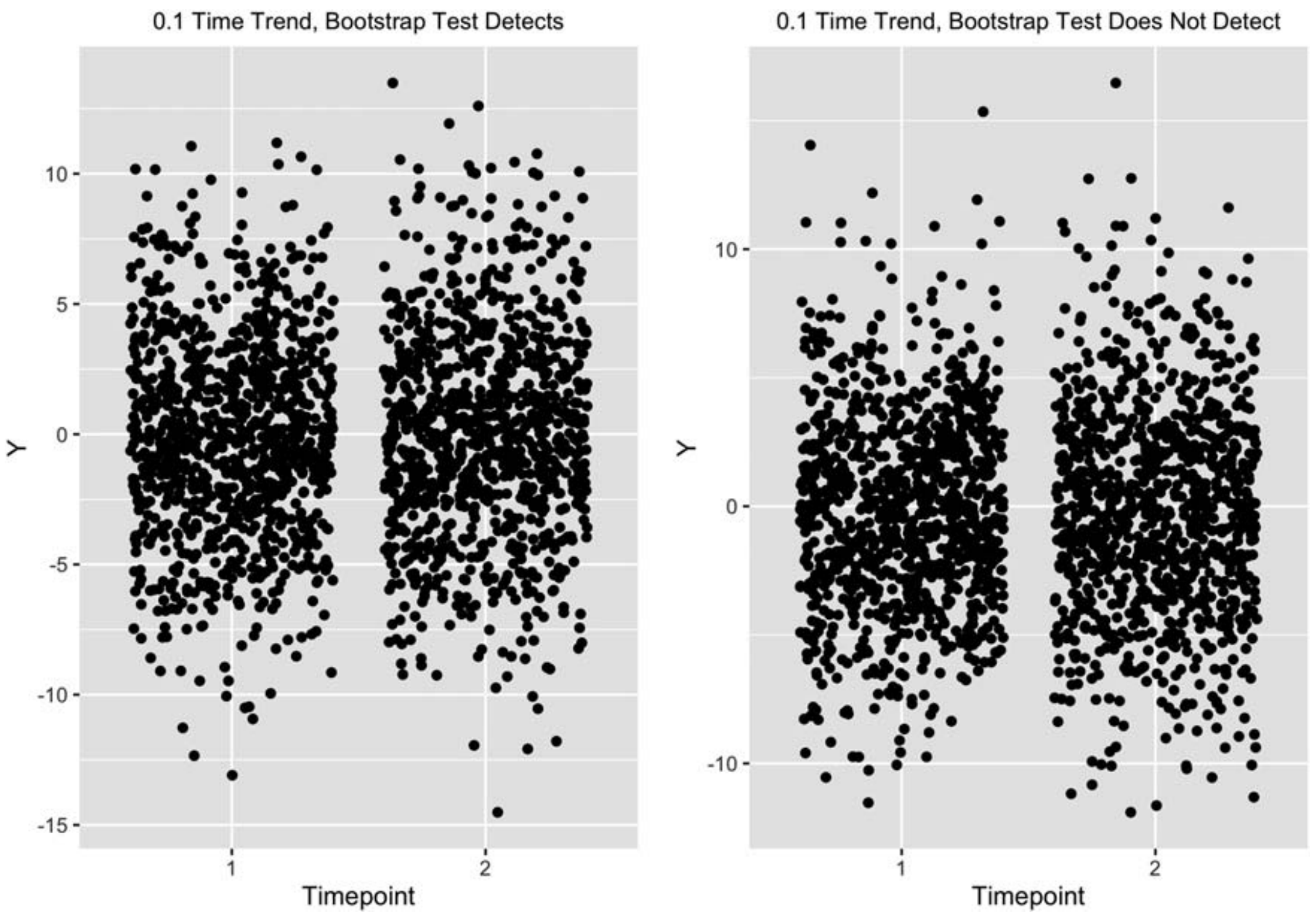

**Figure A1:** Simulated datasets with time trend, $\gamma = 0.1$. The figures show outcome data for a dataset where the timepoint agnosticism test detects the time trend and does not detect the time trend respectively.

$$X_{1,i,t}, X_{2,i,t}, X_{3,i,t_0}, X_{4,i,t_0} \sim N(0,1) \quad \text{for } t = t_0, t_1$$
$$X_{j,i,t_1} = X_{j,i,t_0} + \varepsilon_{j,i} \quad \text{for } j = 3, 4$$
$$\varepsilon_{j,i} \sim N(0, 0.5^2) \quad \text{for } j = 3, 4$$

The outcome variable is a linear combination of the four covariates, a time trend controlled by parameter $\gamma$, and an error term ($\varepsilon_{i,t} \sim N(0,1)$):

$$Y_{i,t} = \log(4)(X_{1,i,t} + X_{4,i,t}) + \log(10)(X_{3,i,t} + X_{4,i,t}) + 1\{t = t_1\}\gamma + \varepsilon_{i,t}$$

We generate data for the setting $\gamma = 0$, which does not include a time varying component, 1,000 times. On each dataset, we perform the test for timepoint agnosticism outlined in this section, with 1-1 matching and bias correction. Our simulated data only contains two timepoints and the sample size is balanced so we choose the first timepoint as $t_0$, our new control group, and the second timepoint as $t_1$, our new treated group. In 0.05 of the simulations the $P$-value is less than 0.05, which shows that type I error is controlled. Now, we add in a time trend where there is an additional term, $\gamma$, added to the second timepoint. For $\gamma = 0.1$ for the second timepoint (resulting in a time trend of 0.1), our simulations result in a $P$-value less than 0.05 in 0.60 of the 1,000 simulations. For $\gamma = 0.25$ for the second timepoint (resulting in a time trend of 0.25), our simulations result in a $P$-value less than 0.05 in all of the 1,000 simulations. Table A4 summarizes these results.

Figure A1 shows two simulated datasets with $\gamma = 0.1$. The test for timepoint agnosticism detects the trend in one of the two datasets. Overall, the simulations show that the test is not a panacea for issues with timepoint agnosticism, sometimes failing to detect small violations. However, it still adds substantial value to the analysis pipeline, detecting moderate violations of timepoint agnosticism not especially obvious to the eye in visualization plots with a very high rate of success.